\documentclass[twocolumn]{aastex631}

\usepackage{amsmath}
\usepackage{color}

\begin{document}

\title{On the optical transients from double white-dwarf mergers}

%\correspondingauthor{August Muench}
%\email{greg.schwarz@aas.org, gus.muench@aas.org}

\author[0000-0002-5438-3460]{M. F. Sousa}
\affiliation{Programa de Pós-Graduação em Física e Astronomia, Universidade Tecnológica Federal do Paraná, Avenida sete de setembro 3165, 80230-901, Curitiba, PR, Brazil}
\affiliation{Divis\~ao de Astrof\'{\i}sica, Instituto Nacional de Pesquisas Espaciais, Avenida dos Astronautas 1758, 12227--010, S\~ao Jos\'e dos Campos, SP, Brazil}
\affiliation{ICRANet-Ferrara, Dip. di Fisica e Scienze della Terra, Universit\`a degli Studi di Ferrara, Via Saragat 1, I--44122 Ferrara, Italy}

\author[0000-0001-9386-1042]{J. G. Coelho}
\affiliation{N\'ucleo de Astrof\'{\i}sica e Cosmologia (Cosmo-Ufes) \& Departamento de F\'isica, Universidade Federal do Esp\'irito Santo, 29075--910, Vit\'oria, ES, Brazil}
\affiliation{Divis\~ao de Astrof\'{\i}sica, Instituto Nacional de Pesquisas Espaciais, Avenida dos Astronautas 1758, 12227--010, S\~ao Jos\'e dos Campos, SP, Brazil}

\author[0000-0003-4418-4289]{J. C. N. de Araujo}
\affiliation{Divis\~ao de Astrof\'{\i}sica, Instituto Nacional de Pesquisas Espaciais, Avenida dos Astronautas 1758, 12227--010, S\~ao Jos\'e dos Campos, SP, Brazil}

\author[0000-0001-6869-0835]{C. Guidorzi}
\affiliation{Department of Physics and Earth Science, 
University of Ferrara, Via Saragat 1, I-44122 Ferrara, Italy}
\affiliation{INFN -- Sezione di Ferrara, Via Saragat 1, 44122 Ferrara, Italy}
\affiliation{INAF -- Osservatorio di Astrofisica e Scienza dello Spazio di Bologna, Via Piero Gobetti 101, 40129 Bologna, Italy}

\author[0000-0003-4904-0014]{J. A. Rueda}
\affiliation{ICRANet, Piazza della Repubblica 10, I-65122 Pescara, Italy}
\affiliation{ICRANet-Ferrara, Dip. di Fisica e Scienze della Terra, Universit\`a degli Studi di Ferrara, Via Saragat 1, I--44122 Ferrara, Italy}
\affiliation{ICRA, Dipartamento di Fisica, Sapienza Universit\`a  di Roma, Piazzale Aldo Moro 5, I-00185 Rome, Italy}
\affiliation{Department of Physics and Earth Science, 
University of Ferrara, Via Saragat 1, I-44122 Ferrara, Italy}
\affiliation{INAF, Istituto di Astrofisica e Planetologia Spaziali, Via Fosso del Cavaliere 100, 00133 Rome, Italy}

%% Note that the \and command from previous versions of AASTeX is now
%% depreciated in this version as it is no longer necessary. AASTeX 
%% automatically takes care of all commas and "and"s between authors names.

%% Mark off the abstract in the ``abstract'' environment. 
\begin{abstract}
Double white-dwarf (DWD) mergers are relevant astrophysical sources expected to produce massive, highly-magnetized WDs, supernovae (SNe) Ia, and neutron stars (NSs). Although they are expected to be numerous sources in the sky, their detection has evaded the most advanced transient surveys. This article characterizes the optical transient expected from DWD mergers in which the central remnant is a stable (sub-Chandrasekhar) WD. We show that the expansion and cooling of the merger's dynamical ejecta lead to an optical emission peaking at $1$--$10$~d post-merger, with luminosities of $10^{40}$--$10^{41}$~erg~s$^{-1}$. We present simulations of the light-curves, spectra, and the color evolution of the transient. We show that these properties, together with the estimated rate of mergers, are consistent with the absence of detection, e.g., by The Zwicky Transient Facility (ZTF). More importantly, we show that the Legacy Survey of Space and Time (LSST) of the Vera C. Rubin Observatory will likely detect a few/several hundred per year, opening a new window to the physics of WDs, NSs, and SN Ia.
\end{abstract}

%% Keywords should appear after the \end{abstract} command. 
%% The AAS Journals now uses Unified Astronomy Thesaurus concepts:
%% https://astrothesaurus.org


%\keywords{Classical Novae (251) --- Ultraviolet astronomy(1736) --- History of astronomy(1868) --- Interdisciplinary astronomy(804)}

%% We recommend that authors also use the natbib \citep
%% and \citet commands to identify citations.  The citations are
%% tied to the reference list via symbolic KEYs. The KEY corresponds
%% to the KEY in the \bibitem in the reference list below. 

%%%%%%%%%%%%%%%%%%%%%%%%%%%%%%%%%%%%%%%%%%%%%%%%%%%%%%
%%%%%%%%%%%%%%%%%%%%%%%%%%%%%%%%%%%%%%%%%%%%%%%%%%%%%%
\section{Introduction} \label{sec:intro}
%%%%%%%%%%%%%%%%%%%%%%%%%%%%%%%%%%%%%%%%%%%%%%%%%%%%%%
%%%%%%%%%%%%%%%%%%%%%%%%%%%%%%%%%%%%%%%%%%%%%%%%%%%%%%

The number of double white dwarfs (DWD) in the Milky Way (MW) merging within a Hubble time has been estimated to be $(5$--$7)\times 10^{-13}$~yr$^{-1}$~$M_\odot^{-1}$ \citep{2017MNRAS.467.1414M, 2018MNRAS.476.2584M}. Using a stellar mass and density of MW-like galaxies of $6.4\times 10^{10}~M_\odot$ and $0.016$~Mpc$^{-3}$ \citep{2001ApJ...556..340K}, it translates into a local cosmic merger rate of ${\cal R}_{\rm DWD} \approx (5$--$7)\times 10^5$~Gpc$^{-3}$~yr$^{-1}$. The above classifies DWD mergers among the most numerous cataclysmic events. 

Three fates of the central remnant of a DWD merger can be envisaged: a fast-rotating (and possibly highly-magnetized) WD, a supernova (SN) of type Ia, or a neutron star (NS). The binary's component masses, the presence (or genesis) of high magnetic fields \citep{2012ApJ...749...25G}, and the rate of mass and angular momentum transfer from a surrounding debris disk are among the critical physical ingredients that determine the central object's fate \citep[see, e.g.,][and references therein]{2018ApJ...857..134B, 2019MNRAS.487..812B}. Based on the above, the relevance of DWDs has been highlighted in various astrophysical scenarios, e.g.:
\begin{itemize}
    \item 
    The double-degenerate scenario
    \citep{1984ApJS...54..335I, 1984ApJ...277..355W} proposes that unstable thermonuclear fusion can be ignited in the central remnant of DWD mergers, leading to one of the most likely explanations of SNe Ia \citep[see, e.g.,][and references therein]{2022ApJ...925...92N}. Indeed, the DWD merger rate is sufficient to explain the rate of SNe Ia, which is about $5$--$8$ times smaller \citep[see, e.g.,][]{2009ApJ...699.2026R, 2018MNRAS.476.2584M}.
    
    \item 
    DWD mergers have been, for a long time, thought to be the main channel leading to the observed WDs with high magnetic fields in the range $10^6$--$10^9$ G \citep{2009A&A...506.1341K, 2015SSRv..191..111F, 2016MNRAS.455.3413K}.
    \item 
    A fraction of DWD mergers can explain the population of massive WDs of $\sim 1 M_\odot$ \citep[see][and references therein]{2018MNRAS.476.2584M, 2020ApJ...891..160C, 2023MNRAS.518.2341K}. See also section \ref{sec:4}.
    \item 
    Interestingly, most of those massive WDs are highly magnetic \citep[see, e.g.,][]{2016MNRAS.455.3413K}. Additionally, LSST will observe more than 150 million WDs at the final depth of its stacked 10-year survey~\cite {2020ApJ...900..139F}.
    \item 
    Indeed, it has been shown that the recently discovered isolated, highly magnetic, rapidly rotating WDs, ZTF J190132.9+145808.7 \citep{2021Natur.595...39C} and SDSS J221141.80+113604.4 \citep{2021ApJ...923L...6K}, could have been formed in DWD mergers \citep[see][for details]{2022ApJ...941...28S}.
     \item 
     Massive, highly magnetized, fast-rotating WDs formed in DWD mergers have been proposed to explain soft gamma repeaters and anomalous X-ray pulsars, i.e., magnetars \citep{2012PASJ...64...56M, 2013ApJ...772L..24R, 2014PASJ...66...14C, 2014ApJ...794...86C, 2016JCAP...05..007M, 2017MNRAS.465.4434C, 2017A&A...599A..87C, 2019ApJ...879...46O, 2020MNRAS.492.5949S, 2020MNRAS.498.4426S, 2020ApJ...895...26B}, fast radio bursts \citep{2013ApJ...776L..39K}, and overluminous SN Ia \citep{2013ApJ...767L..14D, 2022ApJ...926...66D}.
     \item 
     The Laser Interferometer Space Antenna (LISA) expects to detect the gravitational-wave (GW) radiation from many compact (orbital periods shorter than hours), detached DWDs \citep[see, e.g.,][]{2006CQGra..23S.809S, 2022MNRAS.511.5936K, 2022ApJ...940...90C}.
\end{itemize}

Despite the above theoretical and observational richness, additional physical phenomena in DWD mergers have remained unexplored. We aim to characterize them in this article. First of all, given that ${\cal R}_{\rm DWD} \sim (5$--$8){\cal R}_{\rm SN-Ia}$, we must conclude that there is a considerable population of DWD mergers that do not produce SNe~Ia \citep[see, also,][]{ 2020ApJ...891..160C}. This article focuses on such systems, especially those where the central remnant is a massive WD (see section \ref{sec:2}). Section~\ref{sec:3} shows that the dynamical ejecta from DWD mergers produces a fast-rising and fast-declining optical emission, peaking at $\sim 1$ d post-merger, from its cooling driven by the expansion. The energy injected by the central remnant (e.g., by accretion winds and/or pulsar-like emission) is considered. We exemplify such optical transient theoretically and observationally using fiducial model parameters. Section \ref{sec:4} discusses how our findings compare with the known optical transients population. We show the Bright Transient Survey \citep{Perley20} of The Zwicky Transient Facility (ZTF) has not detected/identified any of them. 

Finally, we discuss our main conclusions in Section~\ref{sec:5}, including the consistency of our theoretical predictions with the lack of detections by the ZTF of DWD mergers optical transients. Furthermore, we provide an upper limit for the number of detections expected by the forthcoming Legacy Survey of Space and Time (LSST) of the Vera C. Rubin Observatory. Details on the theoretical modeling of the expected light-curves and spectra are given in the appendix.

%Forthcoming space-based detectors of gravitational waves (GWs) like the Laser Interferometer Space Antenna (LISA), which expects to detect the GW radiation driving the dynamics of compact, detached DWDs  (see, e.g., Refs. \cite{Stroeer2006Sep,Korol2022Apr}).

%%%%%%%%%%%%%%%%%%%%%%%%%%%%%%%%%%%%%%%%%%%%%%%%%%%%%%%%%%%%
%%%%%%%%%%%%%%%%%%%%%%%%%%%%%%%%%%%%%%%%%%%%%%%%%%%%%%%%%%%%
\section{Merging binary and post-merger configuration properties}\label{sec:2} 
%%%%%%%%%%%%%%%%%%%%%%%%%%%%%%%%%%%%%%%%%%%%%%%%%%%%%%%%%%%%
%%%%%%%%%%%%%%%%%%%%%%%%%%%%%%%%%%%%%%%%%%%%%%%%%%%%%%%%%%%%

We are interested in DWD mergers leading to a central remnant that is a stable, sub-Chandrasekhar WD. Given the mass distribution of observed WDs, we expect that sub-Chandra mergers can lead to massive WDs in the $1.0\lesssim M \lesssim 1.4 M_\odot$ range. In principle, such WDs might be fastly rotating with periods $P\gtrsim 0.5$ s \citep[see, e.g.,][]{2013ApJ...762..117B}. Such post-merged WD can avoid exploding as an SN Ia if, during its evolution, its central density remains below some specific value estimated to be a few $10^9$ g cm$^{-3}$ \citep[see, e.g.,][and references therein for details]{2018ApJ...857..134B, 2019MNRAS.487..812B}.

Numerical simulations show that the merger of a DWD, in general, develops a rigidly rotating, central core surrounded by a hot, convective corona with differential rotation and a Keplerian disk that hosts nearly all the mass of the disrupted secondary star \citep{1990ApJ...348..647B,2004A&A...413..257G,2009A&A...500.1193L,2012A&A...542A.117L,2012ApJ...746...62R,2013ApJ...767..164Z,2014MNRAS.438...14D,2018ApJ...857..134B}. These compact-object mergers expel small amounts of mass in the dynamical phase of the merger. \citet{2014MNRAS.438...14D} provided analytic functions that fit the results of their numerical simulations. Concerning the ejected mass, it can be estimated by
\begin{equation}\label{eq:mej}
m_{\rm ej} \approx \frac{0.0001807 M}{-0.01672 + 0.2463 q - 0.6982 q^2 + q^3},
\end{equation}
where $M = m_1 + m_2$ is the total binary mass, and $q \equiv m_2/m_1 \leq 1$ is the binary mass ratio. Equation (\ref{eq:mej}) tells us that, typically, DWD mergers eject $m_{\rm ej} \sim 10^{-3} M_\odot$. Despite this amount of matter being negligible relative to the system mass, we will show that it is responsible for the transient electromagnetic emission in the early post-merger evolution.

%%%%%%%%%%%%%%%%%%%%%%%%%%%%%%%%%%%%%%%%%%%%%%%%%%%%%%%%
%%%%%%%%%%%%%%%%%%%%%%%%%%%%%%%%%%%%%%%%%%%%%%%%%%%%%%%%
\section{Expected light-curves and spectra}\label{sec:3}
%%%%%%%%%%%%%%%%%%%%%%%%%%%%%%%%%%%%%%%%%%%%%%%%%%%%%%%%
%%%%%%%%%%%%%%%%%%%%%%%%%%%%%%%%%%%%%%%%%%%%%%%%%%%%%%%%

We now turn to the results from modeling the emission of the expanding ejecta. As we have recalled, about $10^{-3} M_\odot$ are ejected from the system during the final dynamical phase of the merger. This ejecta expands nearly radially at about the escape velocity, namely, $10^8$--$10^9$ cm s$^{-1}$. In the early post-merger evolution, accretion winds further power the ejecta \citep[see, e.g.,][]{2018ApJ...852..120B, 2019JCAP...03..044R}. Magnetic braking and nuclear reactions can also contribute to the energy budget but to a much lesser extent. In Appendix \ref{app:A}, we present our theoretical model to calculate the thermal evolution of the expanding ejecta subjected to the injection of energy from the central remnant. The model parameters are the ejecta mass ($m_{\rm ej}$), the index defining the radial falloff of the density profile ($m$), the self-similar expansion index ($n$), the initial position and velocity of the innermost ejecta layer ($R_{*,0}$ and $v_{*,0}$), the parameters defining the power injected by the central remnant ($H_0$, $t_c$, and $\delta$), and the optical opacity ($\kappa$). We refer the reader to Appendix \ref{app:A} for technical details. 

\begin{table}
    \centering
    \caption{Parameter values used to model thermal and synchrotron radiation from the expansion of ejected material.}
    \begin{tabular}{lc}
\hline
Parameter & Fiducial Value\\ \hline
$m_{\rm ej}$ $(10^{-3} M_{\odot})$ & 1.00 \\
$n$ & 1.00\\
$m$ & 9.00\\
$R_{*,0}$ ($10^{11}$ cm) & 1.00\\
$v_{*,0}$ ($10^{9}$ cm s$^{-1}$) & 1.00\\
$H_0$ ($10^{46}$ erg s$^{-1}$) & 1.00 \\
$t_c/t_*$ & 1.00 \\
$\delta$ & 1.30 \\
$\kappa$ (cm$^2$ g$^{-1}$) & 0.20 \\
\hline
\end{tabular}
    \label{tab:parameters}
\end{table}

\begin{figure}
\includegraphics[width=\hsize,clip]{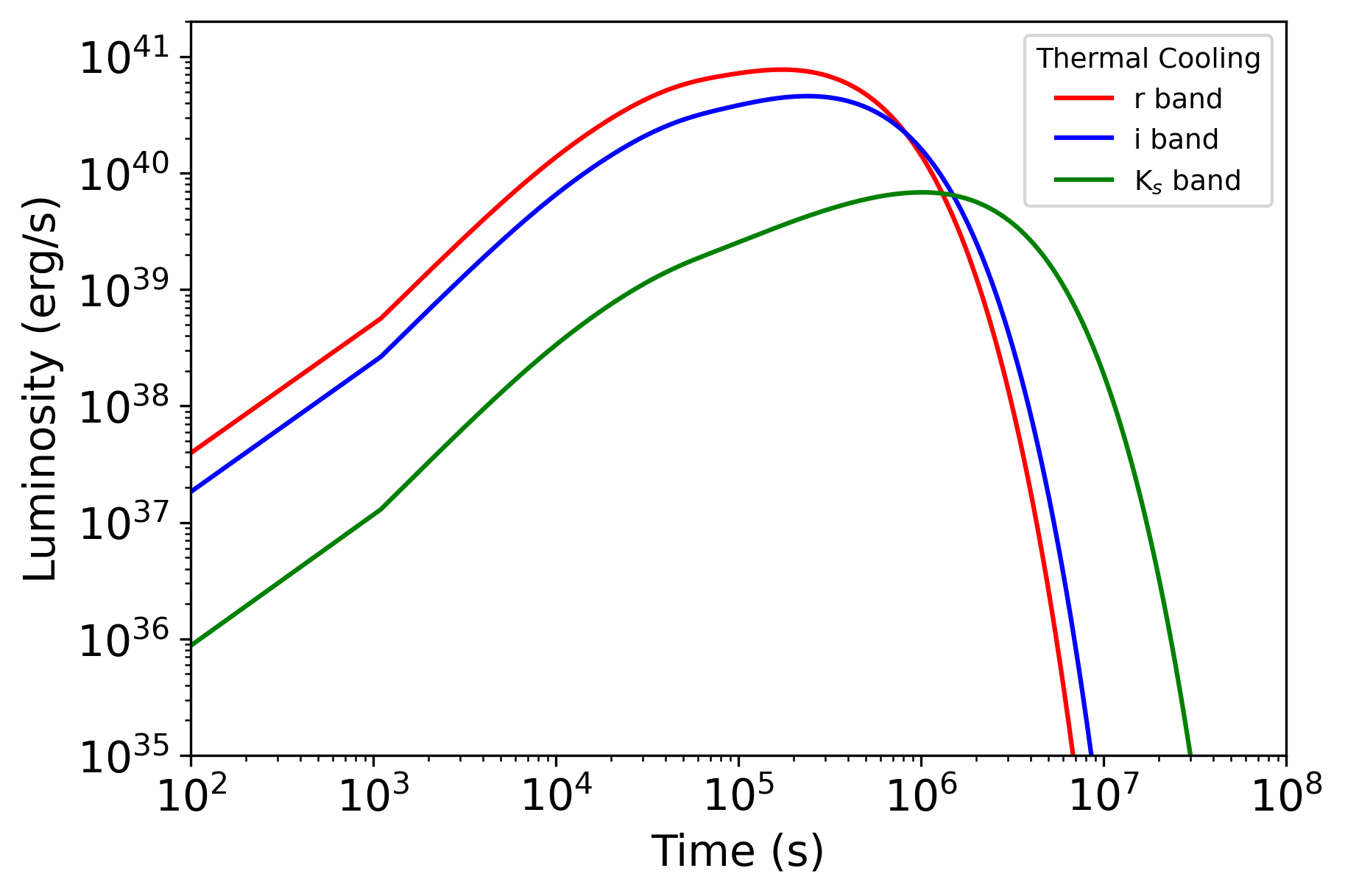}
\caption{Emission from the expanding, cooling ejecta at early times (solid lines) in the visible (r band) and in the infrared ($i$ and $K_s$ bands). We refer to Appendix~\ref{app:A} for details on the theoretical model.} \label{fig:SLumxt}
\end{figure}

Table~\ref{tab:parameters} lists the model parameters and the corresponding fiducial values we adopted to exemplify the model. Figure~\ref{fig:SLumxt} shows the corresponding light-curves (luminosity as a function of time), predicted by the theoretical model in Appendix \ref{app:A}, in the visible (r band) and the infrared ($i$ and $K_s$ bands).

\begin{figure}
\includegraphics[width=\hsize,clip]{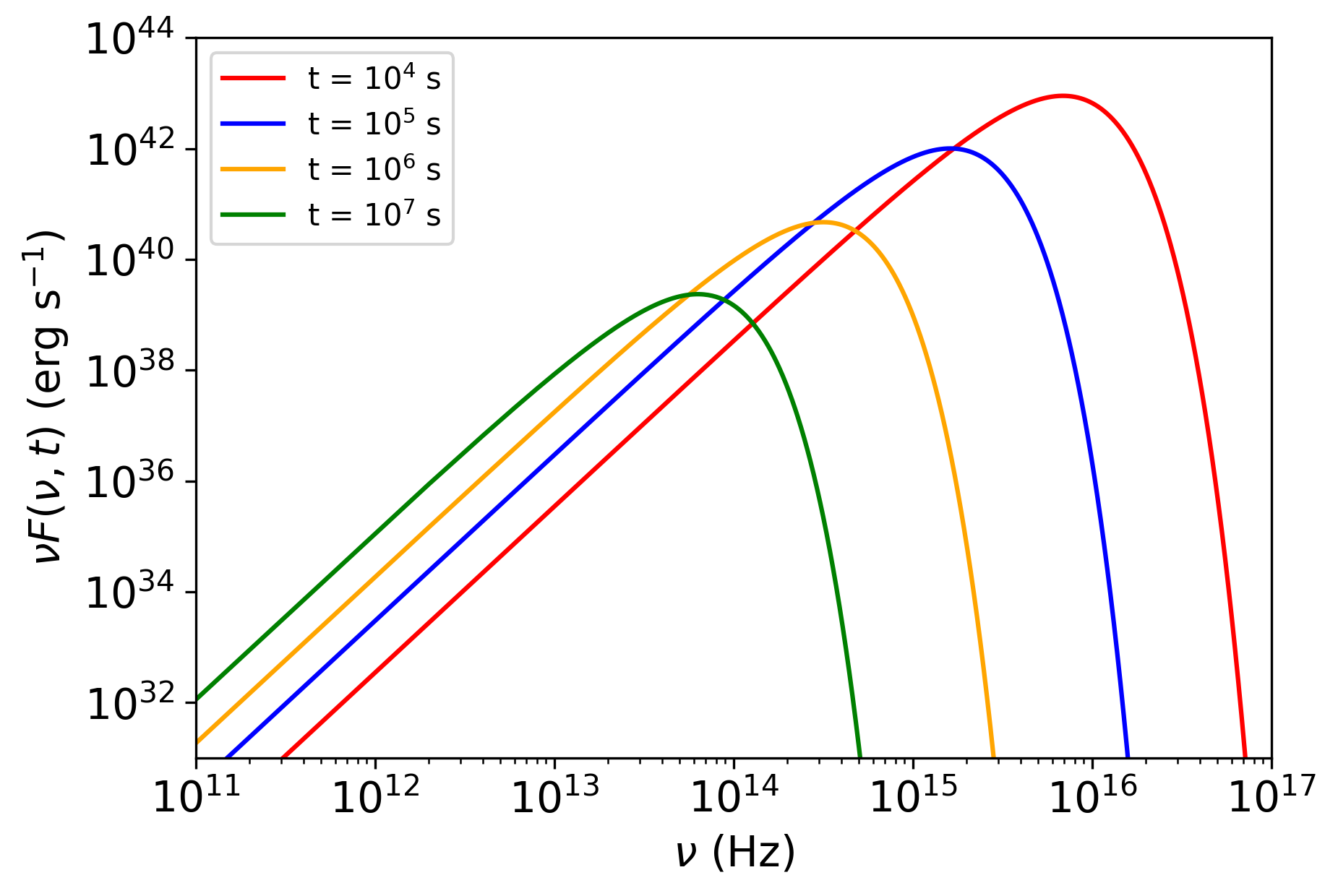}
\caption{Emission spectra from the expanding, cooling ejecta at selected post-merger times. We refer to Appendix~\ref{app:A} for details on the theoretical model} \label{fig:nuFnu}
\end{figure}

From the lightcurves in Fig.~\ref{fig:SLumxt}, we see that the thermal emission due to the expansion of the ejecta peak luminosity is $\sim 10^{40}$--$ 10^{41}$~erg~s$^{-1}$, at about $11$--$12$~d post-merger. The transparency time is $t_{\rm tr,*} \approx 1.55 \times 10^5$~s $\approx 1.79$ d. Figure~\ref{fig:nuFnu} shows the spectra $\nu F(\nu,t)$ at selected times, where $F(\nu,t) = J_{\rm cool}(\nu,t)$ is the spectral density, as given by Eq. (\ref{eq:Jcool}).\\

%%%%%%%%%%%%%%%%%%%%%%%%%%%%%%%%%%%%%%%%%%%%%%%%%%%%%%%%
%%%%%%%%%%%%%%%%%%%%%%%%%%%%%%%%%%%%%%%%%%%%%%%%%%%%%%%%
\section{DWD population, merger rate, and massive WDs}\label{sec:4}
%%%%%%%%%%%%%%%%%%%%%%%%%%%%%%%%%%%%%%%%%%%%%%%%%%%%%%%%
%%%%%%%%%%%%%%%%%%%%%%%%%%%%%%%%%%%%%%%%%%%%%%%%%%%%%%%%

Although the electromagnetic detection of DWDs is a challenging observational task, the increasing quality, sensitivity, and capacity of performing accurate surveys by novel optical observational facilities (e.g., the SDSS, ZTF, Gaia) and the refinement of observational techniques have led to a ten-fold increase in the number of observed DWDs in the Milky Way in the last $20$ years: from around $14$ by $2000$ \citep{2001A&A...365..491N} to about $150$ by $2022$ \citep{2022MNRAS.511.5936K}. That number has already increased \citep[see, e.g.,][]{2023ApJ...950..141K}, also in view of the rapidly growing number of observed WDs in binaries in recent data from the Gaia Mission and ZTF, of which an appreciable percentage are expected to be DWDs \citep[see, e.g.,][]{2023MNRAS.521.1880B, 2023ApJ...950..141K, 2023MNRAS.518.4579P, 2023MNRAS.518.5106J}.

Using population synthesis models that matched the at-the-time number of observed DWDs, i.e., fourteen, in their pioneering work,  \citet{2001A&A...365..491N} estimated the Milky Way hosts about $2.5\times 10^8$ DWDs, and a DWD merger rate $\approx 2.2\times 10^{-2}$ yr$^{-1}$. Up-to-date analyses that match the increasing number of known DWDs have confirmed their estimate. We shall use the estimates by \citet{2018MNRAS.476.2584M}, to which we refer the reader for details. They estimated a DWD merger rate per WD of
${\cal R}_{\rm DWD} = (9.7 \pm 1.1)\times 10^{-12}$ yr$^{-1}$. This estimate can be translated into a DWD merger rate per unit stellar mass by dividing it by the stellar mass to WD number ratio $(15.5 \pm 1.8) M_\odot$ per WD, leading to ${\cal R}_{\rm DWD} \approx (5$--$7)\times 10^{-13}$~yr$^{-1}$~$M_\odot^{-1}$. It is worth noticing that this number agrees with the initial estimate by \citet{2001A&A...365..491N} when multiplied by the Milky Way stellar mass. Assuming a constant star-formation rate over the Milky Way lifetime, these figures imply that $\sim 10\%$ of the Galactic WDs have merged with another WD. Therefore, as discussed in \citet{2018MNRAS.476.2584M}, this inferred fraction of already merged DWDs may explain the high-mass bump in the WD mass function (see also \citealp{2023MNRAS.518.2341K}). However, some massive WDs may have formed from different channels \citep[see, e.g., the case of J004917.14-252556.81 in][]{2023MNRAS.522.2181K}.

The above result agrees with our basic assumption that a considerable fraction of DWD mergers do not lead to SNe Ia but to massive WDs (rapidly rotating and possibly highly magnetic). Therefore, attention must be given to the possibility of establishing the link between observed massive WDs and their possible DWD merger progenitors. The success of this task needs the observational determination of the WD parameters (e.g., mass, radius, rotation period, temperature, and magnetic field strength) and the accurate modeling of the merger and post-merger evolution of the system. Fortunately, there is a growing effort in both directions. Numerical simulations focusing on the merging phase of DWDs started in the 90s and have considerably improved over the years \citep[see, e.g.,][]{1990ApJ...348..647B,2004A&A...413..257G,2009A&A...500.1193L,2012A&A...542A.117L,2012ApJ...746...62R,2013ApJ...767..164Z,2014MNRAS.438...14D,2018ApJ...857..134B}. Theoretical analyses to constrain the physics of the post-merger remnant, to determine its possible fate either as a disrupting explosion (SN Ia), a stable massive WD or gravitational collapse to a NS, including magnetic fields, rotation, and general relativistic effects have also gained interest and been performed in the last decade \citep[see, e.g.,][]{2012MNRAS.427..190S, 2012ApJ...748...35S, 2013ApJ...773..136J, 2013ApJ...776L..39K, 2014MNRAS.438..169B, 2016MNRAS.463.3461S, 2018JCAP...10..006R1, 2018ApJ...857..134B, Shen19, 2022ApJ...925...92N}. Although there is still room for improvements in the merger and post-merger modeling, these works have already allowed us to test the viability of the connection between massive WDs and their possible DWD progenitors on a theoretical basis. For instance, \citet{2022ApJ...941...28S} has positively assessed such a connection for the isolated, highly magnetic, rapidly rotating WDs ZTF J190132.9+145808.7 \citep{2021Natur.595...39C} and SDSS J221141.80+113604.4 \citep{2021ApJ...923L...6K}, leading to the parameters of the possible DWD progenitor, which were found to agree with those of the DWD observed population.

Having set the theoretical and observational basis for the connection between DWD mergers and massive WDs, we next discuss the electromagnetic transient associated with such an astrophysical system, theoretically featured in section \ref{sec:2}, from the observational viewpoint. \\

%%%%%%%%%%%%%%%%%%%%%%%%%%%%%%%%%%%%%%%%%%%%%%%%%%%%%%%%
%%%%%%%%%%%%%%%%%%%%%%%%%%%%%%%%%%%%%%%%%%%%%%%%%%%%%%%%
\section{Observed populations of fast transients}\label{sec:5}
%%%%%%%%%%%%%%%%%%%%%%%%%%%%%%%%%%%%%%%%%%%%%%%%%%%%%%%%
%%%%%%%%%%%%%%%%%%%%%%%%%%%%%%%%%%%%%%%%%%%%%%%%%%%%%%%%

In the last decades, the advent of wide-field, high-cadence surveys led to the discovery of several classes of fast ($t_{\rm rise}\la 10$~d) transients, with luminosities spanning several decades (see \citealt{Pastorello19b} for a review). The so-called `Fast Blue Optical Transients' (FBOTs) are blue, fast-rising, with peak luminosities in the range $-16\gtrsim M_{g,\rm peak} \gtrsim -22$ (e.g., \citealt{Drout14, Tanaka16, Pursiainen18, Tampo20}) and are also referred to as `Rapidly Evolving Transients' or `Fast-Evolving Luminous Transients'. The source AT2018cow (known as `the cow'), at $60$~Mpc, represents the best-studied case of this class. It exhibited some unprecedented characteristics: rise time of a few days; $L_p\sim 4\times10^{44}$~erg~s$^{-1}$; mostly featureless spectra with blackbody temperatures above $10^4$~K during the first $15$~d with large expansion velocities ($\sim 0.1~c$); hard X-ray and variable soft X-ray emission; radio bright with $L_{\nu,p}\sim 4\times10^{28}$~erg~s$^{-1}$~Hz$^{-1}$ at $8.5$~GHz (\citealt{Margutti19_cow, Perley19, Ho19}; see also \citealt{Coppejans20, Ho20, Perley21, Ho23, Matthews23} for the few analogous cases yet observed). These properties suggest that a large amount of radioactive nickel cannot explain the high luminosity, and the relatively short effective diffusion timescale points to a low ejecta mass. In contrast, the long-lived X-ray variability suggests a compact and long-lived inner engine. Owing to their extreme peak luminosity from radio to hard X-rays, these FBOTs are hardly compatible with a DWD merger since the power injected from the central remnant at those times is lower than the observed luminosities.

In parallel, other transients sharing comparably fast rise times ($\sim 12$--$15$~d) but significantly less luminous have also been discovered, with peak luminosities in the gap between novae and supernovae. A class that raised interest is that of so-called calcium-rich transients ($-13\gtrsim M_V\gtrsim -17$; \citealt{Perets10, Kasliwal12, De20}), which exhibit a strong [Ca II] emission in the nebular phase spectra with a high [Ca II ]/[O I] ratio. These share similar photospheric velocities with typical core-collapse Ib/c SNe. Still, their environment is strongly different from the latter since they prefer remote locations in the outskirts of early-type galaxies, even more than type-Ia SNe and short gamma-ray bursts, indicative of a very old progenitor population \citep{Lunnan17}. In this respect, a fraction of these transients could result from mergers of helium and oxygen/neon WDs \citep{Shen19}. The local volumetric rate of Ca-rich, hydrogen-poor transients is estimated to be $\gtrsim 15\%$ of the type-Ia rate \citep{De20}. The variety in peak luminosity and spectroscopic properties probably stems from a heterogeneous class of progenitors.

Some low-luminosity gap transients are still likely to be less energetic SNe. In particular, the so-called intermediate-luminosity red transients (ILRTs; \citealt{Berger09, Bond09}) have a peak luminosity in the range $-12\gtrsim M_V\gtrsim -15$, relatively long rise times and post-peak plateaus which resemble type II-L and II-P SNe. Although there is consensus that the progenitors are $8$--$15~M_{\odot}$ stars in dusty cocoons, eruptive formation of a massive WD or eruptions from binary interactions could contribute to the observed population \citep{Pastorello19b}.

A few gap transients $M_V \gtrsim -13$~mag, characterized by double or even triple-peaked light curves, have been proposed as a scaled-up version of red novae (typically less luminous than $-10$~mag) and, as such, are often referred to as luminous red novae (LRNe; see \citealt{Kulkarni07,Pastorello19a} and references therein). Their photometric evolution is reminiscent of eruptive variables such as V1309 Scorpii, whose final brightening was interpreted as the merger of a contact binary \citep{Tylenda11}.

\begin{figure}[htbp]
\includegraphics[width=\hsize,clip]{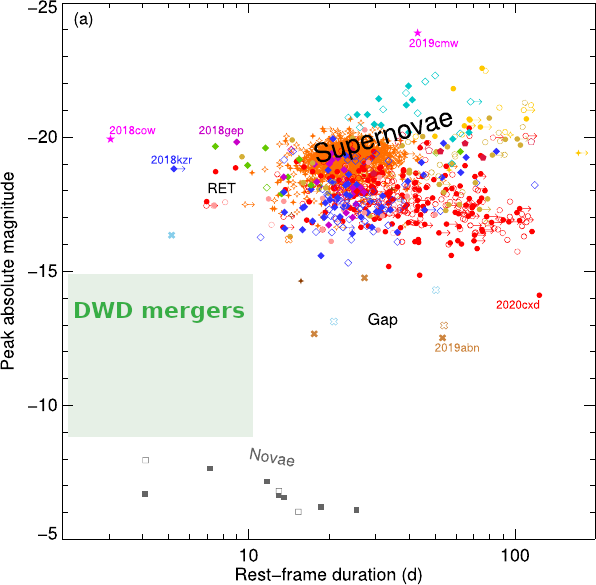}
\caption{Different populations of fast transients observed with the Zwicky Transient Facility Bright Transient Survey \citep{Perley20}. The shaded box highlights where we expect most DWD mergers should lie. Figure adapted from \citet{Perley20}.} \label{fig:obs_transients}
\end{figure}

Figure~\ref{fig:obs_transients} summarizes the zoo of the fast transients as observed with the Zwicky Transient Factory Bright Transient Survey \citep{Perley20} in the peak luminosity--duration plane. We show the region where our predictions on DWD mergers lie: despite the relatively high expected volumetric rate, this region is still poorly explored. Upcoming surveys such as the Legacy Survey of Space and Time (LSST; \citealt{LSST19}) are expected to boost the number of promising candidates for DWD mergers.

%%%%%%%%%%%%%%%%%%%%%%%%%%%%%%%%%%%%%%%%%%%%%%%%%%%%%%%%
%%%%%%%%%%%%%%%%%%%%%%%%%%%%%%%%%%%%%%%%%%%%%%%%%%%%%%%%
\section{Discussion and Conclusions}\label{sec:6}
%%%%%%%%%%%%%%%%%%%%%%%%%%%%%%%%%%%%%%%%%%%%%%%%%%%%%%%%
%%%%%%%%%%%%%%%%%%%%%%%%%%%%%%%%%%%%%%%%%%%%%%%%%%%%%%%%

We have estimated the optical transient from DWD mergers leading to stable, massive, fast-rotating WDs. The emission arises from the cooling down of the dynamical ejecta of the merger, about $10^{-3} M_\odot$, that expands at $10^8$--$10^9$ cm s$^{-1}$. The ejecta is powered by the early activity of the central remnant, mainly fallback accretion (see, e.g., \citealt{2019JCAP...03..044R}, and references therein, and Appendix \ref{app:A} for a comparison of accretion power with nuclear energy and magnetic braking). Inspired by numerical simulations, we assumed spherical expansion. The theoretical model includes a power-law density profile and self-similar expansion. We solve the energy balance equation and determine the ejecta's thermal history (time evolution), estimating its photospheric emission and color evolution.

We have shown that the peak of the optical emission occurs at times $1$--$10$ d, with a luminosity $L_p=10^{40}$--$10^{41}$~erg~s$^{-1}$, for typical parameters expected for these DWD mergers (see Table \ref{tab:parameters}); see Figs. \ref{fig:SLumxt} and \ref{fig:nuFnu} for the light-curves and spectra, respectively. Although our model makes some approximations, we expect it to catch the main physics of these systems robustly. Therefore, further model refinements should not appreciably change the above qualitative and quantitative picture.

With this in mind, we turned to the observational considerations. Indeed, detecting the optical counterpart of DWD mergers would have several relevant consequences in physics and astrophysics. To mention some:

-- It will constrain the fraction of mergers producing SNe Ia, giving crucial hints for the SN Ia-associated physics, e.g., the unstable thermonuclear fusion and detonation \citep[see][and references therein]{2012MNRAS.427..190S, 2012ApJ...748...35S, 2013ApJ...773..136J, 2016MNRAS.463.3461S, 2022ApJ...925...92N}.
 
-- If the rate of mergers leading to SN Ia will turn out lower than the SN Ia observed rate, it would imply the necessity of also having at work the single-degenerate scenario for their explanation \citep[see, e.g.,][]{2004MNRAS.350.1301H}. 

-- It will alert facilities on ground and space to look for associated emissions at higher energies, e.g., in the X- and gamma-rays, constraining the physics of the central remnant such as magnetic fields and rotation \citep[see, e.g.,][]{2013ApJ...773..136J, 2013ApJ...776L..39K, 2014MNRAS.438..169B, 2018JCAP...10..006R1, 2018ApJ...857..134B}.

-- It will confirm DWD mergers as the formation channel of massive, fast-rotating WDs;

-- At late post-merger times, the central WD might be observed accompanied by a debris disk \citep[see, e.g.,][and references therein]{2013MNRAS.431.2778K, 2013ApJ...772L..24R, 2018ApJ...857..134B, 2022ApJ...925...92N}. Thus, it will be interesting to compare forthcoming advanced infrared-optical-UV survey estimates of the rate of WDs with debris disk \citep{2020ApJ...900..139F} and the DWD merger rate estimates.

-- It will constrain the physics of the gravitational collapse of WDs into NSs while simultaneously possibly confirming DWD mergers as a formation channel of NSs.

Thus, in section \ref{sec:4}, we checked whether current observational facilities could have observed such optical transients. We compare and contrast the model predictions with the emergence population of optical transients in the literature. Our analysis showed that the optical transients from DWD mergers presented here do not match the observed features of FBOTs, fast-evolving luminous transients (i.e., cow-like objects), and calcium-rich transients. A plot of the peak absolute magnitude as a function of the rest-frame time duration for the transients detected by the ZTD Bright Transient Survey (see Fig. \ref{fig:obs_transients}), highlighting the region the DWD merger optical transients should occupy, reveals overwhelmingly the above result. 

Therefore, no optical transient from DWD mergers has ever been detected. Does this result agree with the model prediction? The limiting magnitude for detection by ZTF is $m_{\rm ZTF, lim}=19$~mag \citep{Perley20}. Assuming the peak luminosity $L_p=10^{40}$~erg~s$^{-1}$, this turns into a detection horizon $d_{\rm ZTF,lim}\sim 11$~Mpc. Using an expected volumetric rate for DWD mergers of $4\times 10^5$~Gpc$^{-3}$~yr$^{-1}$ (see section \ref{sec:intro}), the upper limit to the expected number of events by ZTF is $\sim 2$, considering the duty cycle of the survey. This result is consistent with our findings and the expectation that not all DWD mergers produce stable WDs: a fraction should lead to SNe Ia and another to NSs as central remnants. 

We can apply the same kind of calculation to LSST, for which 5-$\sigma$ limiting magnitudes for single exposures in filters $g$ and $r$ (the same considered for ZTF in Fig.~\ref{fig:obs_transients}) are $24.5$ and $24.0$, respectively\footnote{\url{https://www.lsst.org/scientists/keynumbers}}. Under these favorable conditions, the detection horizon becomes $d_{\rm LSST,lim}\sim 110$--$140$~Mpc, corresponding to a gain by $\sim10^3$ in the expected detection rate. 

The above analysis brings us to one of the main conclusions: in the transition from ZTF to LSST, the electromagnetic (optical) counterparts of DWD mergers will finally become observable, likely a few/several hundred per year, opening a new window to the physics of WDs, NSs, and SN Ia.

%% Also note that the akcnowlodgment environment does not support long amounts of text. If you have a lot of people and institutions to acknowledge, do not use this command. Instead, create a new \section{Acknowledgments}.

%\begin{acknowledgments}
%M.F.S. thanks CAPES-PrInt (88887.351889/2019-00) and Fundação Araucária for the financial support. J.G.C. is grateful for the support of FAPES (1020/2022, 1081/2022, 976/2022, 332/2023), CNPq (311758/2021-5), and FAPESP (2021/01089-1). J.C.N.A. thanks CNPq (307803/2022-8) for partial financial support.
%\end{acknowledgments}

\vspace{0.8cm}
{\it{Acknowledgments:} M.F.S. thanks CAPES-PrInt (88887.351889/2019-00) and Fundação Araucária for the financial support. J.G.C. is grateful for the support of FAPES (1020/2022, 1081/2022, 976/2022, 332/2023), CNPq (311758/2021-5), and FAPESP (2021/01089-1). J.C.N.A. thanks CNPq (307803/2022-8) for partial financial support.}

%\vspace{5mm}
%\facilities{HST(STIS), Swift(XRT and UVOT), AAVSO, CTIO:1.3m,
%CTIO:1.5m,CXO}

%\software{astropy \citep{2013A&A...558A..33A,2018AJ....156..123A},  
%          Cloudy \citep{2013RMxAA..49..137F}, 
%          Source Extractor \citep{1996A&AS..117..393B}
%          }

\appendix

%%%%%%%%%%%%%%%%%%%%%%%%%%%%%%%%%%%%%%%%%%%%%%%%%%%%%%
%%%%%%%%%%%%%%%%%%%%%%%%%%%%%%%%%%%%%%%%%%%%%%%%%%%%%%
\section{Emission from the cooling of the expanding ejecta}\label{app:A}
%%%%%%%%%%%%%%%%%%%%%%%%%%%%%%%%%%%%%%%%%%%%%%%%%%%%%%
%%%%%%%%%%%%%%%%%%%%%%%%%%%%%%%%%%%%%%%%%%%%%%%%%%%%%%

For modeling the thermal emission of the expanding ejecta, we must consider that the layers reach transparency at different times in a non-homogeneous distribution of matter. The present model generalizes the model presented in \citet{2019JCAP...03..044R}. Numerical simulations show that the ejected matter expands nearly radially, so we consider a spherically symmetric distribution. The ejecta extends at radii $r_i \in [R_*, R_{\rm max}]$, with corresponding velocities $v_i \in [v_*,v_{\rm max}]$, in self-similar expansion
\begin{equation}\label{eq:radius}
    r_i(t) = r_{i,0} \hat{t}^n,\quad v_i(t) = n \frac{r_i(t)}{t} = v_{i,0} \hat{t}^{n-1},
\end{equation}
where $\hat{t} \equiv t/t_*$, being $t_* \equiv n R_{*,0}/v_{*,0}$ the characteristic expansion timescale, which is the same for all layers given the condition of self-similarity. Here, $r_{i,0}$ and $v_{i,0}$ are the initial radius and velocity of the layer. The case $n=1$ corresponds to a uniform expansion.

The density at the position $r=r_i$ is given by
\begin{equation}\label{eq:rhoej}
    \rho(r_i) = \frac{(3-m)}{4\pi} \frac{m_{\rm ej}}{R_{*,0}^3} \frac{(R_*/r_i)^m}{\left[\left(\frac{R_{\rm max}}{R_*}\right)^{3-m} -1\right]}\hat{t}^{-3 n},
\end{equation}
where $m_{\rm ej}$ is the total mass of the ejecta, and $m$ is a positive constant. The distribution and time evolution given by Eq. (\ref{eq:rhoej}) ensure that at any time, the total mass of the ejecta, i.e., the volume integral of the density, equals $m_{\rm ej}$.

We divide the ejecta into $N$ shells defined by the $N+1$ radii 
\begin{equation}\label{eq:radii}
    r_{i,0} = R_{*,0} + i \frac{(R_{\rm max,0}-R_{*,0})}{N},\quad i=0,1,...,N,
\end{equation}
so the width and mass of each shell are, respectively, $\Delta r =  (R_{\rm max}-R_{*})/N$, and
\begin{equation}\label{eq:mi}
    m_i = \int_{r_i}^{r_{i+1}} 4\pi r^2 \rho(r) dr \approx \frac{4\pi}{m-3} r_i^3 \rho(r_i) \left [ 1 - \left ( 1 + \frac{\Delta r}{r_{i}} \right )^{3-m} \right ],
\end{equation}
so given the decreasing density with distance, the inner layers are more massive than the outer layers. The number of shells to be used must satisfy the constraint that the sum of the shells' mass gives the total ejecta mass, i.e.
\begin{equation}\label{eq:mjsum}
    \sum_{j=1}^{N} m_j = m_{\rm ej}.
\end{equation}
We have introduced the discrete index $j=i+1$ to differentiate the counting of the shells from the counting of radii given by Eq. (\ref{eq:radii}). In this work, we use $N=100$ shells, ensuring that Eq. (\ref{eq:mjsum}) is satisfied with $99\%$ of accuracy. 

Under the assumption that the shells do not interact with each other, we can estimate the evolution of the $i$-th shell from the energy conservation equation
\begin{equation}\label{eq:energybalance}
\dot{E}_i = -P_i\,\dot{V}_i - L_{{\rm cool},i} + H_{{\rm inj},i},
\end{equation}
where $V_{i} = (4\pi/3)r_{i}^3$, $E_i$, and $P_i$ are the volume, energy, and pressure of the shell, while $H_{{\rm inj},i}$ is the power injected from the central remnant that is thermalized in the shell, and
\begin{equation}\label{eq:Lcoolabs}
     L_{{\rm cool},i} \approx \frac{c E_i}{r_i (1+\tau_{{\rm opt},i})},
\end{equation}
is the bolometric luminosity radiated by the shell, being $\tau_{\rm opt,i}$ the optical depth. 

Assuming a spatially constant gray opacity throughout the ejecta, the optical depth of the radiation emitted by the $i$-th layer is given by
\begin{equation}\label{eq:taui}
    \tau_{\rm opt,i} = \int_{\infty}^{r_i} \kappa \rho(r) dr = \int_{R_{\rm max}}^{r_i} \kappa \rho(r) dr = \tau_{i,0} \hat{t}^{-2n}, \qquad 
    \tau_{i,0} \equiv \frac{m-3}{m-1} \frac{\kappa m_{\rm ej}}{4\pi R_{*,0}^2}\frac{\left[ \left(\frac{R_*}{r_i}\right)^{m-1} - \left(\frac{R_*}{r_{\rm max}}\right)^{m-1} \right]}{\left[1- \left(\frac{R_*}{R_{\rm max}}\right)^{m-3}\right]},
\end{equation}
where we have used Eq. (\ref{eq:rhoej}), and $\kappa$ is the opacity. 

We adopt a radiation-dominated equation of state for the ejecta, so at every position, $E_i \approx 3 P_i\,V_i$.  The power injected into the ejecta originates from the newborn central WD \citep{2019JCAP...03..044R}. This energy is absorbed and thermalized, becoming a heating source for the expanding matter. The power-law decreasing density, Eq. (\ref{eq:rhoej}), suggests that the inner the layer, the more radiation it should absorb. To account for this effect, we weigh the heating source for each shell using the mass fraction, i.e.
\begin{equation}\label{eq:Hi}
    H_{{\rm inj},i} = \frac{m_i}{m_{\rm ej}} H_{\rm inj},
\end{equation}
where $m_i$ is the shell's mass, and adopt the following form for the heating source 
\begin{equation}\label{eq:Hinj}
    H_{\rm inj} = H_0 \left(1+\frac{t}{t_c}\right)^{-\delta},
\end{equation}
where $H_0$ and $\delta$ are model parameters. This function can model the power injected by the pulsar's spindown by magnetic braking and accretion winds. We expect that power released from accretion dominates the early times. For instance, for fallback accretion parameters $H_0 = 10^{46}$ erg s$^{-1}$, $\delta = 1.3$, and $t_c = t_*$ \citep{2019JCAP...03..044R}, with $t_* = 10^2$~s, we obtain $H_{\rm inj} \approx 2 \times 10^{43}$ erg s$^{-1}$ at $t=10^4$ s. We can set an upper limit on the energy injected from nuclear reactions, e.g., by nickel decay, assuming nickel amounts to the entire ejecta mass, i.e., $M_{\rm Ni} = m_{\rm ej} \sim 10^{-3} M_\odot$. In that case, reactions would inject $L_{\rm Ni} = 3.9\times 10^{10} M_{\rm Ni}\,e^{-t/(8.8 d)}\sim 10^{41}$ erg s$^{-1}$ by that time, which is still much smaller than the expected power injected by fallback accretion. Likewise, magnetic braking leads to negligible energy injection from rotational energy loss at early times. For example, a WD with a dipole magnetic field of strength $B_d = 10^9$ G, radius $R=10^8$ cm, initial rotation period $P_0 = 10$ s ($\Omega_0 = 2\pi/P_0 \approx 0.6$ rad s$^{-1}$), and moment of inertia $I=10^{49}$ g cm$^2$, has a characteristic magnetic braking timescale $t_{\rm sd} = T/L_{\rm sd} \approx 2 \times 10^7$ yr, where $T = (1/2)I \Omega_0^2$ is the initial rotational energy and $L_{\rm d} = (2/3) B_d^2 R^6 \Omega_0^4/c^3 $ the initial spin-down power due to magnetic dipole braking. Therefore, at times $t \lesssim t_{\rm sd} \approx 6\times 10^{14}$ s, the spin-down power is $L_{\rm d} \approx 4\times 10^{33}$ erg s$^{-1}$. From the above, we safely assume a single source of injection power, modeled by Eq. (\ref{eq:Hinj}), bearing in mind that other power inputs could be considered but which should have a negligible effect in the early post-merger transient.

The position of the shell that reaches transparency gives the photospheric radius at a time $t$. Namely, the shell's position with optical depth $\tau_{\rm opt, i}[r_i(t)] = 1$. Using Eq. (\ref{eq:taui}), we obtain 
\begin{equation}\label{eq:Rph}
    R_{\rm ph} = \frac{R_{{\rm max},0} \hat{t}^n }{\left[1+\frac{m-1}{m-3}\frac{4\pi R_{*,0}^2}{\kappa m_{\rm ej}} \frac{ \left[1- \left(\frac{R_*}{R_{\rm max}}\right)^{m-3} \right]}{\left(\frac{R_*}{R_{\rm max}}\right)^{m-1}} \hat{t}^{2 n}\right]^{\frac{1}{m-1}}}.
\end{equation}

Equation (\ref{eq:Rph}) shows that when the entire ejecta is optically thick, $R_{\rm ph} = R_{\rm max}$. Then, the transparency reaches the inner shells to the instant over which $R_{\rm ph} = R_*$, at $t=t_{\rm tr,*}$, when the entire ejecta has become transparent. The time $t_{\rm tr,*}$ is found from the condition $\tau_{\rm opt,*}[R_*(t_{\rm tr,*})] = 1$, and is given by
\begin{equation}\label{eq:ttr}
    \hat{t}_{\rm tr,*} = \left\{\frac{m-3}{m-1} \frac{\kappa m_{\rm ej}}{4\pi R_{*,0}^2}\left(\frac{R_*}{R_{\rm max}}\right)^{m-1}\frac{\left[ \left(\frac{R_{\rm max}}{R_*}\right)^{m-1} - 1 \right]}{\left[1- \left(\frac{R_*}{R_{\rm max}}\right)^{m-3} \right]}\right\}^{\frac{1}{2 n}}.
\end{equation}

At $t<t_{\rm tr,*}$, the photospheric radius evolves as $R_{\rm ph} \propto t^{\frac{n (m-3)}{m-1}}$, while at later times, $R_{\rm ph} \propto t^n$.

The sum of the luminosity of the shells gives the bolometric luminosity
\begin{equation}
    L_{\rm bol} = \sum_{j=1}^{N} L_{{\rm cool},j},
\end{equation}
so the effective temperature of the blackbody emission, $T_s$, can be obtained from the Stefan–Boltzmann law, i.e. 
\begin{equation}\label{eq:Teff}
T_s = \left(\frac{L_{\rm bol}}{4\pi R_{\rm ph}^2 \sigma}\right)^{1/4},
\end{equation}
where $\sigma$ is the Stefan-Boltzmann constant. The power per unit frequency, per unit area, is given by Planck's spectrum
\begin{equation}\label{eq:Bnu}
    B_\nu = \frac{2\pi h \nu^3}{c^2} \frac{1}{e^{\frac{h\nu}{k_B T_s}} - 1},
\end{equation}
where $\nu$ is the radiation frequency, $h$ and $k_B$ are the Planck and Boltzmann constants. Therefore, the spectral density (power per unit frequency) given by the thermal cooling at a frequency $\nu$ is
\begin{equation}\label{eq:Jcool}
   J_{\rm cool}(\nu,t) = 4\pi R_{\rm ph}^2(t) B_\nu(\nu,t),
\end{equation}
and the luminosity radiated in the frequency range $[\nu_1,\nu_2]$ can be then obtained as
\begin{equation}\label{eq:Lnucool}
    L_{\rm cool}(\nu_1,\nu_2; t) = \int_{\nu_1}^{\nu_2} J_{\rm cool}(\nu,t)d\nu.
\end{equation}

The parameter $v_{\rm max,0}$ has no appreciable effect in the evolution, so it can not be constrained from the data. This happens because most of the mass is concentrated in the innermost layers, so they dominate the thermal evolution. For self-consistency of the model, we have set $v_{\rm max,0} = 2 v_{*,0}$ (so $R_{\rm max,0} = 2 R_{*,0}$). As for the initial value of the internal energy of the shells, $E_i(t_0)$, we have set them to the initial kinetic energy of each layer, $E_i = (1/2) m_i v_i(t_0)^2$. 

\bibliography{sample631,references}

\begin{thebibliography}{}
\expandafter\ifx\csname natexlab\endcsname\relax\def\natexlab#1{#1}\fi
\providecommand{\url}[1]{\href{#1}{#1}}
\providecommand{\dodoi}[1]{doi:~\href{http://doi.org/#1}{\nolinkurl{#1}}}
\providecommand{\doeprint}[1]{\href{http://ascl.net/#1}{\nolinkurl{http://ascl.net/#1}}}
\providecommand{\doarXiv}[1]{\href{https://arxiv.org/abs/#1}{\nolinkurl{https://arxiv.org/abs/#1}}}

\bibitem[{{Becerra} {et~al.}(2019){Becerra}, {Boshkayev}, {Rueda}, \&
  {Ruffini}}]{2019MNRAS.487..812B}
{Becerra}, L., {Boshkayev}, K., {Rueda}, J.~A., \& {Ruffini}, R. 2019, \mnras,
  487, 812, \dodoi{10.1093/mnras/stz1394}

\bibitem[{{Becerra} {et~al.}(2018{\natexlab{a}}){Becerra}, {Guzzo},
  {Rossi-Torres}, {Rueda}, {Ruffini}, \& {Uribe}}]{2018ApJ...852..120B}
{Becerra}, L., {Guzzo}, M.~M., {Rossi-Torres}, F., {et~al.} 2018{\natexlab{a}},
  \apj, 852, 120, \dodoi{10.3847/1538-4357/aaa296}

\bibitem[{{Becerra} {et~al.}(2018{\natexlab{b}}){Becerra}, {Rueda},
  {Lor{\'e}n-Aguilar}, \& {Garc{\'{\i}}a-Berro}}]{2018ApJ...857..134B}
{Becerra}, L., {Rueda}, J.~A., {Lor{\'e}n-Aguilar}, P., \&
  {Garc{\'{\i}}a-Berro}, E. 2018{\natexlab{b}}, \apj, 857, 134,
  \dodoi{10.3847/1538-4357/aabc12}

\bibitem[{{Beloborodov}(2014)}]{2014MNRAS.438..169B}
{Beloborodov}, A.~M. 2014, \mnras, 438, 169, \dodoi{10.1093/mnras/stt2140}

\bibitem[{{Benz} {et~al.}(1990){Benz}, {Cameron}, {Press}, \&
  {Bowers}}]{1990ApJ...348..647B}
{Benz}, W., {Cameron}, A.~G.~W., {Press}, W.~H., \& {Bowers}, R.~L. 1990, \apj,
  348, 647, \dodoi{10.1086/168273}

\bibitem[{{Berger} {et~al.}(2009){Berger}, {Soderberg}, {Chevalier},
  {Fransson}, {Foley}, {Leonard}, {Debes}, {Diamond-Stanic}, {Dupree}, {Ivans},
  {Simmerer}, {Thompson}, \& {Tremonti}}]{Berger09}
{Berger}, E., {Soderberg}, A.~M., {Chevalier}, R.~A., {et~al.} 2009, \apj, 699,
  1850, \dodoi{10.1088/0004-637X/699/2/1850}

\bibitem[{{Bond} {et~al.}(2009){Bond}, {Bedin}, {Bonanos}, {Humphreys},
  {Monard}, {Prieto}, \& {Walter}}]{Bond09}
{Bond}, H.~E., {Bedin}, L.~R., {Bonanos}, A.~Z., {et~al.} 2009, \apjl, 695,
  L154, \dodoi{10.1088/0004-637X/695/2/L154}

\bibitem[{{Borges} {et~al.}(2020){Borges}, {Rodrigues}, {Coelho}, {Malheiro},
  \& {Castro}}]{2020ApJ...895...26B}
{Borges}, S.~V., {Rodrigues}, C.~V., {Coelho}, J.~G., {Malheiro}, M., \&
  {Castro}, M. 2020, \apj, 895, 26, \dodoi{10.3847/1538-4357/ab8add}

\bibitem[{{Boshkayev} {et~al.}(2013){Boshkayev}, {Rueda}, {Ruffini}, \&
  {Siutsou}}]{2013ApJ...762..117B}
{Boshkayev}, K., {Rueda}, J.~A., {Ruffini}, R., \& {Siutsou}, I. 2013, \apj,
  762, 117, \dodoi{10.1088/0004-637X/762/2/117}

\bibitem[{{Brown} {et~al.}(2023){Brown}, {Parsons}, {van Roestel},
  {Rebassa-Mansergas}, {Breedt}, {Dhillon}, {Dyer}, {Green}, {Kerry},
  {Littlefair}, {Marsh}, {Munday}, {Pelisoli}, {Sahman}, \&
  {Wild}}]{2023MNRAS.521.1880B}
{Brown}, A.~J., {Parsons}, S.~G., {van Roestel}, J., {et~al.} 2023, \mnras,
  521, 1880, \dodoi{10.1093/mnras/stad612}

\bibitem[{{C{\'a}ceres} {et~al.}(2017){C{\'a}ceres}, {de Carvalho}, {Coelho},
  {de Lima}, \& {Rueda}}]{2017MNRAS.465.4434C}
{C{\'a}ceres}, D.~L., {de Carvalho}, S.~M., {Coelho}, J.~G., {de Lima},
  R.~C.~R., \& {Rueda}, J.~A. 2017, \mnras, 465, 4434,
  \dodoi{10.1093/mnras/stw3047}

\bibitem[{{Caiazzo} {et~al.}(2021){Caiazzo}, {Burdge}, {Fuller}, {Heyl},
  {Kulkarni}, {Prince}, {Richer}, {Schwab}, {Andreoni}, {Bellm}, {Drake},
  {Duev}, {Graham}, {Helou}, {Mahabal}, {Masci}, {Smith}, \&
  {Soumagnac}}]{2021Natur.595...39C}
{Caiazzo}, I., {Burdge}, K.~B., {Fuller}, J., {et~al.} 2021, \nat, 595, 39,
  \dodoi{10.1038/s41586-021-03615-y}

\bibitem[{{Carvalho} {et~al.}(2022){Carvalho}, {Anjos}, {Coelho}, {Lobato},
  {Malheiro}, {Marinho}, {Rodriguez}, {Rueda}, \&
  {Ruffini}}]{2022ApJ...940...90C}
{Carvalho}, G.~A., {Anjos}, R.~C.~d., {Coelho}, J.~G., {et~al.} 2022, \apj,
  940, 90, \dodoi{10.3847/1538-4357/ac9841}

\bibitem[{{Cheng} {et~al.}(2020){Cheng}, {Cummings}, {M{\'e}nard}, \&
  {Toonen}}]{2020ApJ...891..160C}
{Cheng}, S., {Cummings}, J.~D., {M{\'e}nard}, B., \& {Toonen}, S. 2020, \apj,
  891, 160, \dodoi{10.3847/1538-4357/ab733c}

\bibitem[{{Coelho} {et~al.}(2017){Coelho}, {C{\'a}ceres}, {de Lima},
  {Malheiro}, {Rueda}, \& {Ruffini}}]{2017A&A...599A..87C}
{Coelho}, J.~G., {C{\'a}ceres}, D.~L., {de Lima}, R.~C.~R., {et~al.} 2017,
  \aap, 599, A87, \dodoi{10.1051/0004-6361/201629521}

\bibitem[{{Coelho} \& {Malheiro}(2014)}]{2014PASJ...66...14C}
{Coelho}, J.~G., \& {Malheiro}, M. 2014, \pasj, 66, 14,
  \dodoi{10.1093/pasj/pst014}

\bibitem[{{Coelho} {et~al.}(2014){Coelho}, {Marinho}, {Malheiro}, {Negreiros},
  {C{\'a}ceres}, {Rueda}, \& {Ruffini}}]{2014ApJ...794...86C}
{Coelho}, J.~G., {Marinho}, R.~M., {Malheiro}, M., {et~al.} 2014, \apj, 794,
  86, \dodoi{10.1088/0004-637X/794/1/86}

\bibitem[{{Coppejans} {et~al.}(2020){Coppejans}, {Margutti}, {Terreran},
  {Nayana}, {Coughlin}, {Laskar}, {Alexander}, {Bietenholz}, {Caprioli},
  {Chandra}, {Drout}, {Frederiks}, {Frohmaier}, {Hurley}, {Kochanek},
  {MacLeod}, {Meisner}, {Nugent}, {Ridnaia}, {Sand}, {Svinkin}, {Ward}, {Yang},
  {Baldeschi}, {Chilingarian}, {Dong}, {Esquivia}, {Fong}, {Guidorzi},
  {Lundqvist}, {Milisavljevic}, {Paterson}, {Reichart}, {Shappee}, {Stroh},
  {Valenti}, {Zauderer}, \& {Zhang}}]{Coppejans20}
{Coppejans}, D.~L., {Margutti}, R., {Terreran}, G., {et~al.} 2020, \apjl, 895,
  L23, \dodoi{10.3847/2041-8213/ab8cc7}

\bibitem[{{Dan} {et~al.}(2014){Dan}, {Rosswog}, {Br{\"u}ggen}, \&
  {Podsiadlowski}}]{2014MNRAS.438...14D}
{Dan}, M., {Rosswog}, S., {Br{\"u}ggen}, M., \& {Podsiadlowski}, P. 2014,
  \mnras, 438, 14, \dodoi{10.1093/mnras/stt1766}

\bibitem[{{Das} {et~al.}(2013){Das}, {Mukhopadhyay}, \&
  {Rao}}]{2013ApJ...767L..14D}
{Das}, U., {Mukhopadhyay}, B., \& {Rao}, A.~R. 2013, \apjl, 767, L14,
  \dodoi{10.1088/2041-8205/767/1/L14}

\bibitem[{{De} {et~al.}(2020){De}, {Kasliwal}, {Tzanidakis}, {Fremling},
  {Adams}, {Aloisi}, {Andreoni}, {Bagdasaryan}, {Bellm}, {Bildsten},
  {Cannella}, {Cook}, {Delacroix}, {Drake}, {Duev}, {Dugas}, {Frederick},
  {Gal-Yam}, {Goldstein}, {Golkhou}, {Graham}, {Hale}, {Hankins}, {Helou},
  {Ho}, {Irani}, {Jencson}, {Kaplan}, {Kaye}, {Kulkarni}, {Kupfer}, {Laher},
  {Leadbeater}, {Lunnan}, {Masci}, {Miller}, {Neill}, {Ofek}, {Perley},
  {Polin}, {Prince}, {Quataert}, {Reiley}, {Riddle}, {Rusholme}, {Sharma},
  {Shupe}, {Sollerman}, {Tartaglia}, {Walters}, {Yan}, \& {Yao}}]{De20}
{De}, K., {Kasliwal}, M.~M., {Tzanidakis}, A., {et~al.} 2020, \apj, 905, 58,
  \dodoi{10.3847/1538-4357/abb45c}

\bibitem[{{Deb} {et~al.}(2022){Deb}, {Mukhopadhyay}, \&
  {Weber}}]{2022ApJ...926...66D}
{Deb}, D., {Mukhopadhyay}, B., \& {Weber}, F. 2022, \apj, 926, 66,
  \dodoi{10.3847/1538-4357/ac410b}

\bibitem[{{Drout} {et~al.}(2014){Drout}, {Chornock}, {Soderberg}, {Sanders},
  {McKinnon}, {Rest}, {Foley}, {Milisavljevic}, {Margutti}, {Berger},
  {Calkins}, {Fong}, {Gezari}, {Huber}, {Kankare}, {Kirshner}, {Leibler},
  {Lunnan}, {Mattila}, {Marion}, {Narayan}, {Riess}, {Roth}, {Scolnic},
  {Smartt}, {Tonry}, {Burgett}, {Chambers}, {Hodapp}, {Jedicke}, {Kaiser},
  {Magnier}, {Metcalfe}, {Morgan}, {Price}, \& {Waters}}]{Drout14}
{Drout}, M.~R., {Chornock}, R., {Soderberg}, A.~M., {et~al.} 2014, \apj, 794,
  23, \dodoi{10.1088/0004-637X/794/1/23}

\bibitem[{{Fantin} {et~al.}(2020){Fantin}, {C{\^o}t{\'e}}, \&
  {McConnachie}}]{2020ApJ...900..139F}
{Fantin}, N.~J., {C{\^o}t{\'e}}, P., \& {McConnachie}, A.~W. 2020, \apj, 900,
  139, \dodoi{10.3847/1538-4357/aba270}

\bibitem[{{Ferrario} {et~al.}(2015){Ferrario}, {de Martino}, \&
  {G{\"a}nsicke}}]{2015SSRv..191..111F}
{Ferrario}, L., {de Martino}, D., \& {G{\"a}nsicke}, B.~T. 2015, \ssr, 191,
  111, \dodoi{10.1007/s11214-015-0152-0}

\bibitem[{{Garc{\'{\i}}a-Berro} {et~al.}(2012){Garc{\'{\i}}a-Berro},
  {Lor{\'e}n-Aguilar}, {Aznar-Sigu{\'a}n}, {Torres}, {Camacho}, {Althaus},
  {C{\'o}rsico}, {K{\"u}lebi}, \& {Isern}}]{2012ApJ...749...25G}
{Garc{\'{\i}}a-Berro}, E., {Lor{\'e}n-Aguilar}, P., {Aznar-Sigu{\'a}n}, G.,
  {et~al.} 2012, \apj, 749, 25, \dodoi{10.1088/0004-637X/749/1/25}

\bibitem[{{Guerrero} {et~al.}(2004){Guerrero}, {Garc{\'{\i}}a-Berro}, \&
  {Isern}}]{2004A&A...413..257G}
{Guerrero}, J., {Garc{\'{\i}}a-Berro}, E., \& {Isern}, J. 2004, \aap, 413, 257,
  \dodoi{10.1051/0004-6361:20031504}

\bibitem[{{Han} \& {Podsiadlowski}(2004)}]{2004MNRAS.350.1301H}
{Han}, Z., \& {Podsiadlowski}, P. 2004, \mnras, 350, 1301,
  \dodoi{10.1111/j.1365-2966.2004.07713.x}

\bibitem[{{Ho} {et~al.}(2019){Ho}, {Phinney}, {Ravi}, {Kulkarni}, {Petitpas},
  {Emonts}, {Bhalerao}, {Blundell}, {Cenko}, {Dobie}, {Howie}, {Kamraj},
  {Kasliwal}, {Murphy}, {Perley}, {Sridharan}, \& {Yoon}}]{Ho19}
{Ho}, A. Y.~Q., {Phinney}, E.~S., {Ravi}, V., {et~al.} 2019, \apj, 871, 73,
  \dodoi{10.3847/1538-4357/aaf473}

\bibitem[{{Ho} {et~al.}(2020){Ho}, {Perley}, {Kulkarni}, {Dong}, {De},
  {Chandra}, {Andreoni}, {Bellm}, {Burdge}, {Coughlin}, {Dekany}, {Feeney},
  {Frederiks}, {Fremling}, {Golkhou}, {Graham}, {Hale}, {Helou}, {Horesh},
  {Kasliwal}, {Laher}, {Masci}, {Miller}, {Porter}, {Ridnaia}, {Rusholme},
  {Shupe}, {Soumagnac}, \& {Svinkin}}]{Ho20}
{Ho}, A. Y.~Q., {Perley}, D.~A., {Kulkarni}, S.~R., {et~al.} 2020, \apj, 895,
  49, \dodoi{10.3847/1538-4357/ab8bcf}

\bibitem[{Ho {et~al.}(2023)Ho, Perley, Gal-Yam, Lunnan, Sollerman, Schulze,
  Das, Dobie, Yao, Fremling, Adams, Anand, Andreoni, Bellm, Bruch, Burdge,
  Castro-Tirado, Dahiwale, De, Dekany, Drake, Duev, Graham, Helou, Kaplan,
  Karambelkar, Kasliwal, Kool, Kulkarni, Mahabal, Medford, Miller, Nordin,
  Ofek, Petitpas, Riddle, Sharma, Smith, Stewart, Taggart, Tartaglia,
  Tzanidakis, \& Winters}]{Ho23}
Ho, A. Y.~Q., Perley, D.~A., Gal-Yam, A., {et~al.} 2023, \apj, 949, 120,
  \dodoi{10.3847/1538-4357/acc533}

\bibitem[{{Iben} \& {Tutukov}(1984)}]{1984ApJS...54..335I}
{Iben}, I., J., \& {Tutukov}, A.~V. 1984, \apjs, 54, 335,
  \dodoi{10.1086/190932}

\bibitem[{{Ivezi{\'c}} {et~al.}(2019){Ivezi{\'c}}, {Kahn}, {Tyson}, {Abel},
  {Acosta}, {Allsman}, {Alonso}, {AlSayyad}, {Anderson}, {Andrew}, {Angel},
  {Angeli}, {Ansari}, {Antilogus}, {Araujo}, {Armstrong}, {Arndt}, {Astier},
  {Aubourg}, {Auza}, {Axelrod}, {Bard}, {Barr}, {Barrau}, {Bartlett}, {Bauer},
  {Bauman}, {Baumont}, {Bechtol}, {Bechtol}, {Becker}, {Becla}, {Beldica},
  {Bellavia}, {Bianco}, {Biswas}, {Blanc}, {Blazek}, {Blandford}, {Bloom},
  {Bogart}, {Bond}, {Booth}, {Borgland}, {Borne}, {Bosch}, {Boutigny},
  {Brackett}, {Bradshaw}, {Brandt}, {Brown}, {Bullock}, {Burchat}, {Burke},
  {Cagnoli}, {Calabrese}, {Callahan}, {Callen}, {Carlin}, {Carlson},
  {Chandrasekharan}, {Charles-Emerson}, {Chesley}, {Cheu}, {Chiang}, {Chiang},
  {Chirino}, {Chow}, {Ciardi}, {Claver}, {Cohen-Tanugi}, {Cockrum}, {Coles},
  {Connolly}, {Cook}, {Cooray}, {Covey}, {Cribbs}, {Cui}, {Cutri}, {Daly},
  {Daniel}, {Daruich}, {Daubard}, {Daues}, {Dawson}, {Delgado}, {Dellapenna},
  {de Peyster}, {de Val-Borro}, {Digel}, {Doherty}, {Dubois},
  {Dubois-Felsmann}, {Durech}, {Economou}, {Eifler}, {Eracleous}, {Emmons},
  {Fausti Neto}, {Ferguson}, {Figueroa}, {Fisher-Levine}, {Focke}, {Foss},
  {Frank}, {Freemon}, {Gangler}, {Gawiser}, {Geary}, {Gee}, {Geha}, {Gessner},
  {Gibson}, {Gilmore}, {Glanzman}, {Glick}, {Goldina}, {Goldstein}, {Goodenow},
  {Graham}, {Gressler}, {Gris}, {Guy}, {Guyonnet}, {Haller}, {Harris},
  {Hascall}, {Haupt}, {Hernandez}, {Herrmann}, {Hileman}, {Hoblitt}, {Hodgson},
  {Hogan}, {Howard}, {Huang}, {Huffer}, {Ingraham}, {Innes}, {Jacoby}, {Jain},
  {Jammes}, {Jee}, {Jenness}, {Jernigan}, {Jevremovi{\'c}}, {Johns}, {Johnson},
  {Johnson}, {Jones}, {Juramy-Gilles}, {Juri{\'c}}, {Kalirai}, {Kallivayalil},
  {Kalmbach}, {Kantor}, {Karst}, {Kasliwal}, {Kelly}, {Kessler}, {Kinnison},
  {Kirkby}, {Knox}, {Kotov}, {Krabbendam}, {Krughoff}, {Kub{\'a}nek},
  {Kuczewski}, {Kulkarni}, {Ku}, {Kurita}, {Lage}, {Lambert}, {Lange},
  {Langton}, {Le Guillou}, {Levine}, {Liang}, {Lim}, {Lintott}, {Long},
  {Lopez}, {Lotz}, {Lupton}, {Lust}, {MacArthur}, {Mahabal}, {Mandelbaum},
  {Markiewicz}, {Marsh}, {Marshall}, {Marshall}, {May}, {McKercher}, {McQueen},
  {Meyers}, {Migliore}, {Miller}, {Mills}, {Miraval}, {Moeyens}, {Moolekamp},
  {Monet}, {Moniez}, {Monkewitz}, {Montgomery}, {Morrison}, {Mueller},
  {Muller}, {Mu{\~n}oz Arancibia}, {Neill}, {Newbry}, {Nief}, {Nomerotski},
  {Nordby}, {O'Connor}, {Oliver}, {Olivier}, {Olsen}, {O'Mullane}, {Ortiz},
  {Osier}, {Owen}, {Pain}, {Palecek}, {Parejko}, {Parsons}, {Pease},
  {Peterson}, {Peterson}, {Petravick}, {Libby Petrick}, {Petry},
  {Pierfederici}, {Pietrowicz}, {Pike}, {Pinto}, {Plante}, {Plate}, {Plutchak},
  {Price}, {Prouza}, {Radeka}, {Rajagopal}, {Rasmussen}, {Regnault}, {Reil},
  {Reiss}, {Reuter}, {Ridgway}, {Riot}, {Ritz}, {Robinson}, {Roby}, {Roodman},
  {Rosing}, {Roucelle}, {Rumore}, {Russo}, {Saha}, {Sassolas}, {Schalk},
  {Schellart}, {Schindler}, {Schmidt}, {Schneider}, {Schneider}, {Schoening},
  {Schumacher}, {Schwamb}, {Sebag}, {Selvy}, {Sembroski}, {Seppala}, {Serio},
  {Serrano}, {Shaw}, {Shipsey}, {Sick}, {Silvestri}, {Slater}, {Smith},
  {Smith}, {Sobhani}, {Soldahl}, {Storrie-Lombardi}, {Stover}, {Strauss},
  {Street}, {Stubbs}, {Sullivan}, {Sweeney}, {Swinbank}, {Szalay}, {Takacs},
  {Tether}, {Thaler}, {Thayer}, {Thomas}, {Thornton}, {Thukral}, {Tice},
  {Trilling}, {Turri}, {Van Berg}, {Vanden Berk}, {Vetter}, {Virieux},
  {Vucina}, {Wahl}, {Walkowicz}, {Walsh}, {Walter}, {Wang}, {Wang}, {Warner},
  {Wiecha}, {Willman}, {Winters}, {Wittman}, {Wolff}, {Wood-Vasey}, {Wu},
  {Xin}, {Yoachim}, \& {Zhan}}]{LSST19}
{Ivezi{\'c}}, {\v{Z}}., {Kahn}, S.~M., {Tyson}, J.~A., {et~al.} 2019, \apj,
  873, 111, \dodoi{10.3847/1538-4357/ab042c}

\bibitem[{{Ji} {et~al.}(2013){Ji}, {Fisher}, {Garc{\'\i}a-Berro}, {Tzeferacos},
  {Jordan}, {Lee}, {Lor{\'e}n-Aguilar}, {Cremer}, \&
  {Behrends}}]{2013ApJ...773..136J}
{Ji}, S., {Fisher}, R.~T., {Garc{\'\i}a-Berro}, E., {et~al.} 2013, \apj, 773,
  136, \dodoi{10.1088/0004-637X/773/2/136}

\bibitem[{{Jim{\'e}nez-Esteban} {et~al.}(2023){Jim{\'e}nez-Esteban}, {Torres},
  {Rebassa-Mansergas}, {Cruz}, {Murillo-Ojeda}, {Solano}, {Rodrigo}, \&
  {Camisassa}}]{2023MNRAS.518.5106J}
{Jim{\'e}nez-Esteban}, F.~M., {Torres}, S., {Rebassa-Mansergas}, A., {et~al.}
  2023, \mnras, 518, 5106, \dodoi{10.1093/mnras/stac3382}

\bibitem[{{Kalogera} {et~al.}(2001){Kalogera}, {Narayan}, {Spergel}, \&
  {Taylor}}]{2001ApJ...556..340K}
{Kalogera}, V., {Narayan}, R., {Spergel}, D.~N., \& {Taylor}, J.~H. 2001, \apj,
  556, 340, \dodoi{10.1086/321583}

\bibitem[{{Kashiyama} {et~al.}(2013){Kashiyama}, {Ioka}, \&
  {M{\'e}sz{\'a}ros}}]{2013ApJ...776L..39K}
{Kashiyama}, K., {Ioka}, K., \& {M{\'e}sz{\'a}ros}, P. 2013, \apjl, 776, L39,
  \dodoi{10.1088/2041-8205/776/2/L39}

\bibitem[{{Kasliwal} {et~al.}(2012){Kasliwal}, {Kulkarni}, {Gal-Yam}, {Nugent},
  {Sullivan}, {Bildsten}, {Yaron}, {Perets}, {Arcavi}, {Ben-Ami}, {Bhalerao},
  {Bloom}, {Cenko}, {Filippenko}, {Frail}, {Ganeshalingam}, {Horesh}, {Howell},
  {Law}, {Leonard}, {Li}, {Ofek}, {Polishook}, {Poznanski}, {Quimby},
  {Silverman}, {Sternberg}, \& {Xu}}]{Kasliwal12}
{Kasliwal}, M.~M., {Kulkarni}, S.~R., {Gal-Yam}, A., {et~al.} 2012, \apj, 755,
  161, \dodoi{10.1088/0004-637X/755/2/161}

\bibitem[{{Kepler} {et~al.}(2016){Kepler}, {Pelisoli}, {Koester}, {Ourique},
  {Romero}, {Reindl}, {Kleinman}, {Eisenstein}, {Valois}, \&
  {Amaral}}]{2016MNRAS.455.3413K}
{Kepler}, S.~O., {Pelisoli}, I., {Koester}, D., {et~al.} 2016, \mnras, 455,
  3413, \dodoi{10.1093/mnras/stv2526}

\bibitem[{{Kilic} {et~al.}(2023{\natexlab{a}}){Kilic}, {C{\'o}rsico}, {Moss},
  {Jewett}, {De Ger{\'o}nimo}, \& {Althaus}}]{2023MNRAS.522.2181K}
{Kilic}, M., {C{\'o}rsico}, A.~H., {Moss}, A.~G., {et~al.} 2023{\natexlab{a}},
  \mnras, 522, 2181, \dodoi{10.1093/mnras/stad1113}

\bibitem[{{Kilic} {et~al.}(2021){Kilic}, {Kosakowski}, {Moss}, {Bergeron}, \&
  {Conly}}]{2021ApJ...923L...6K}
{Kilic}, M., {Kosakowski}, A., {Moss}, A.~G., {Bergeron}, P., \& {Conly}, A.~A.
  2021, \apjl, 923, L6, \dodoi{10.3847/2041-8213/ac3b60}

\bibitem[{{Kilic} {et~al.}(2023{\natexlab{b}}){Kilic}, {Moss}, {Kosakowski},
  {Bergeron}, {Conly}, {Brown}, {Toonen}, {Williams}, \&
  {Dufour}}]{2023MNRAS.518.2341K}
{Kilic}, M., {Moss}, A.~G., {Kosakowski}, A., {et~al.} 2023{\natexlab{b}},
  \mnras, 518, 2341, \dodoi{10.1093/mnras/stac3182}

\bibitem[{{Korol} {et~al.}(2022){Korol}, {Hallakoun}, {Toonen}, \&
  {Karnesis}}]{2022MNRAS.511.5936K}
{Korol}, V., {Hallakoun}, N., {Toonen}, S., \& {Karnesis}, N. 2022, \mnras,
  511, 5936, \dodoi{10.1093/mnras/stac415}

\bibitem[{{Kosakowski} {et~al.}(2023){Kosakowski}, {Brown}, {Kilic}, {Kupfer},
  {B{\'e}dard}, {Gianninas}, {Ag{\"u}eros}, \&
  {Barrientos}}]{2023ApJ...950..141K}
{Kosakowski}, A., {Brown}, W.~R., {Kilic}, M., {et~al.} 2023, \apj, 950, 141,
  \dodoi{10.3847/1538-4357/acd187}

\bibitem[{{K{\"u}lebi} {et~al.}(2013){K{\"u}lebi}, {Ek{\c{s}}i},
  {Lor{\'e}n-Aguilar}, {Isern}, \& {Garc{\'\i}a-Berro}}]{2013MNRAS.431.2778K}
{K{\"u}lebi}, B., {Ek{\c{s}}i}, K.~Y., {Lor{\'e}n-Aguilar}, P., {Isern}, J., \&
  {Garc{\'\i}a-Berro}, E. 2013, \mnras, 431, 2778, \dodoi{10.1093/mnras/stt374}

\bibitem[{{K{\"u}lebi} {et~al.}(2009){K{\"u}lebi}, {Jordan}, {Euchner},
  {G{\"a}nsicke}, \& {Hirsch}}]{2009A&A...506.1341K}
{K{\"u}lebi}, B., {Jordan}, S., {Euchner}, F., {G{\"a}nsicke}, B.~T., \&
  {Hirsch}, H. 2009, \aap, 506, 1341, \dodoi{10.1051/0004-6361/200912570}

\bibitem[{{Kulkarni} {et~al.}(2007){Kulkarni}, {Ofek}, {Rau}, {Cenko},
  {Soderberg}, {Fox}, {Gal-Yam}, {Capak}, {Moon}, {Li}, {Filippenko}, {Egami},
  {Kartaltepe}, \& {Sanders}}]{Kulkarni07}
{Kulkarni}, S.~R., {Ofek}, E.~O., {Rau}, A., {et~al.} 2007, \nat, 447, 458,
  \dodoi{10.1038/nature05822}

\bibitem[{{Longland} {et~al.}(2012){Longland}, {Lor{\'e}n-Aguilar}, {Jos{\'e}},
  {Garc{\'{\i}}a-Berro}, \& {Althaus}}]{2012A&A...542A.117L}
{Longland}, R., {Lor{\'e}n-Aguilar}, P., {Jos{\'e}}, J., {Garc{\'{\i}}a-Berro},
  E., \& {Althaus}, L.~G. 2012, \aap, 542, A117,
  \dodoi{10.1051/0004-6361/201219289}

\bibitem[{{Lor{\'e}n-Aguilar} {et~al.}(2009){Lor{\'e}n-Aguilar}, {Isern}, \&
  {Garc{\'{\i}}a-Berro}}]{2009A&A...500.1193L}
{Lor{\'e}n-Aguilar}, P., {Isern}, J., \& {Garc{\'{\i}}a-Berro}, E. 2009, \aap,
  500, 1193, \dodoi{10.1051/0004-6361/200811060}

\bibitem[{{Lunnan} {et~al.}(2017){Lunnan}, {Kasliwal}, {Cao}, {Hangard},
  {Yaron}, {Parrent}, {McCully}, {Gal-Yam}, {Mulchaey}, {Ben-Ami},
  {Filippenko}, {Fremling}, {Fruchter}, {Howell}, {Koda}, {Kupfer}, {Kulkarni},
  {Laher}, {Masci}, {Nugent}, {Ofek}, {Yagi}, \& {Yan}}]{Lunnan17}
{Lunnan}, R., {Kasliwal}, M.~M., {Cao}, Y., {et~al.} 2017, \apj, 836, 60,
  \dodoi{10.3847/1538-4357/836/1/60}

\bibitem[{{Malheiro} {et~al.}(2012){Malheiro}, {Rueda}, \&
  {Ruffini}}]{2012PASJ...64...56M}
{Malheiro}, M., {Rueda}, J.~A., \& {Ruffini}, R. 2012, \pasj, 64, 56,
  \dodoi{10.1093/pasj/64.3.56}

\bibitem[{{Maoz} \& {Hallakoun}(2017)}]{2017MNRAS.467.1414M}
{Maoz}, D., \& {Hallakoun}, N. 2017, \mnras, 467, 1414,
  \dodoi{10.1093/mnras/stx102}

\bibitem[{{Maoz} {et~al.}(2018){Maoz}, {Hallakoun}, \&
  {Badenes}}]{2018MNRAS.476.2584M}
{Maoz}, D., {Hallakoun}, N., \& {Badenes}, C. 2018, \mnras, 476, 2584,
  \dodoi{10.1093/mnras/sty339}

\bibitem[{{Margutti} {et~al.}(2019){Margutti}, {Metzger}, {Chornock}, {Vurm},
  {Roth}, {Grefenstette}, {Savchenko}, {Cartier}, {Steiner}, {Terreran},
  {Margalit}, {Migliori}, {Milisavljevic}, {Alexander}, {Bietenholz},
  {Blanchard}, {Bozzo}, {Brethauer}, {Chilingarian}, {Coppejans}, {Ducci},
  {Ferrigno}, {Fong}, {G{\"o}tz}, {Guidorzi}, {Hajela}, {Hurley}, {Kuulkers},
  {Laurent}, {Mereghetti}, {Nicholl}, {Patnaude}, {Ubertini}, {Banovetz},
  {Bartel}, {Berger}, {Coughlin}, {Eftekhari}, {Frederiks}, {Kozlova},
  {Laskar}, {Svinkin}, {Drout}, {MacFadyen}, \& {Paterson}}]{Margutti19_cow}
{Margutti}, R., {Metzger}, B.~D., {Chornock}, R., {et~al.} 2019, \apj, 872, 18,
  \dodoi{10.3847/1538-4357/aafa01}

\bibitem[{{Matthews} {et~al.}(2023){Matthews}, {Margutti}, {Metzger},
  {Milisavljevic}, {Migliori}, {Laskar}, {Brethauer}, {Berger}, {Chornock},
  {Drout}, \& {Ramirez-Ruiz}}]{Matthews23}
{Matthews}, D.~J., {Margutti}, R., {Metzger}, B.~D., {et~al.} 2023, arXiv
  e-prints, arXiv:2306.01114, \dodoi{10.48550/arXiv.2306.01114}

\bibitem[{{Mukhopadhyay} \& {Rao}(2016)}]{2016JCAP...05..007M}
{Mukhopadhyay}, B., \& {Rao}, A.~R. 2016, \jcap, 2016, 007,
  \dodoi{10.1088/1475-7516/2016/05/007}

\bibitem[{{Nelemans} {et~al.}(2001){Nelemans}, {Yungelson}, {Portegies Zwart},
  \& {Verbunt}}]{2001A&A...365..491N}
{Nelemans}, G., {Yungelson}, L.~R., {Portegies Zwart}, S.~F., \& {Verbunt}, F.
  2001, \aap, 365, 491, \dodoi{10.1051/0004-6361:20000147}

\bibitem[{{Neopane} {et~al.}(2022){Neopane}, {Bhargava}, {Fisher}, {Ferrari},
  {Yoshida}, {Toonen}, \& {Bravo}}]{2022ApJ...925...92N}
{Neopane}, S., {Bhargava}, K., {Fisher}, R., {et~al.} 2022, \apj, 925, 92,
  \dodoi{10.3847/1538-4357/ac3b52}

\bibitem[{{Otoniel} {et~al.}(2019){Otoniel}, {Franzon}, {Carvalho}, {Malheiro},
  {Schramm}, \& {Weber}}]{2019ApJ...879...46O}
{Otoniel}, E., {Franzon}, B., {Carvalho}, G.~A., {et~al.} 2019, \apj, 879, 46,
  \dodoi{10.3847/1538-4357/ab24d1}

\bibitem[{{Parsons} {et~al.}(2023){Parsons}, {Hernandez}, {Toloza},
  {Zorotovic}, {Schreiber}, {G{\"a}nsicke}, {Lagos}, {Raddi},
  {Rebassa-Mansergas}, {Ren}, \& {Koester}}]{2023MNRAS.518.4579P}
{Parsons}, S.~G., {Hernandez}, M.~S., {Toloza}, O., {et~al.} 2023, \mnras, 518,
  4579, \dodoi{10.1093/mnras/stac3368}

\bibitem[{{Pastorello} \& {Fraser}(2019)}]{Pastorello19b}
{Pastorello}, A., \& {Fraser}, M. 2019, Nature Astronomy, 3, 676,
  \dodoi{10.1038/s41550-019-0809-9}

\bibitem[{{Pastorello} {et~al.}(2019){Pastorello}, {Mason}, {Taubenberger},
  {Fraser}, {Cortini}, {Tomasella}, {Botticella}, {Elias-Rosa}, {Kotak},
  {Smartt}, {Benetti}, {Cappellaro}, {Turatto}, {Tartaglia}, {Djorgovski},
  {Drake}, {Berton}, {Briganti}, {Brimacombe}, {Bufano}, {Cai}, {Chen},
  {Christensen}, {Ciabattari}, {Congiu}, {Dimai}, {Inserra}, {Kankare},
  {Magill}, {Maguire}, {Martinelli}, {Morales-Garoffolo}, {Ochner}, {Pignata},
  {Reguitti}, {Sollerman}, {Spiro}, {Terreran}, \& {Wright}}]{Pastorello19a}
{Pastorello}, A., {Mason}, E., {Taubenberger}, S., {et~al.} 2019, \aap, 630,
  A75, \dodoi{10.1051/0004-6361/201935999}

\bibitem[{{Perets} {et~al.}(2010){Perets}, {Gal-Yam}, {Mazzali}, {Arnett},
  {Kagan}, {Filippenko}, {Li}, {Arcavi}, {Cenko}, {Fox}, {Leonard}, {Moon},
  {Sand}, {Soderberg}, {Anderson}, {James}, {Foley}, {Ganeshalingam}, {Ofek},
  {Bildsten}, {Nelemans}, {Shen}, {Weinberg}, {Metzger}, {Piro}, {Quataert},
  {Kiewe}, \& {Poznanski}}]{Perets10}
{Perets}, H.~B., {Gal-Yam}, A., {Mazzali}, P.~A., {et~al.} 2010, \nat, 465,
  322, \dodoi{10.1038/nature09056}

\bibitem[{{Perley} {et~al.}(2019){Perley}, {Mazzali}, {Yan}, {Cenko}, {Gezari},
  {Taggart}, {Blagorodnova}, {Fremling}, {Mockler}, {Singh}, {Tominaga},
  {Tanaka}, {Watson}, {Ahumada}, {Anupama}, {Ashall}, {Becerra}, {Bersier},
  {Bhalerao}, {Bloom}, {Butler}, {Copperwheat}, {Coughlin}, {De}, {Drake},
  {Duev}, {Frederick}, {Gonz{\'a}lez}, {Goobar}, {Heida}, {Ho}, {Horst},
  {Hung}, {Itoh}, {Jencson}, {Kasliwal}, {Kawai}, {Khanam}, {Kulkarni},
  {Kumar}, {Kumar}, {Kutyrev}, {Lee}, {Maeda}, {Mahabal}, {Murata}, {Neill},
  {Ngeow}, {Penprase}, {Pian}, {Quimby}, {Ramirez-Ruiz}, {Richer},
  {Rom{\'a}n-Z{\'u}{\~n}iga}, {Sahu}, {Srivastav}, {Socia}, {Sollerman},
  {Tachibana}, {Taddia}, {Tinyanont}, {Troja}, {Ward}, {Wee}, \&
  {Yu}}]{Perley19}
{Perley}, D.~A., {Mazzali}, P.~A., {Yan}, L., {et~al.} 2019, \mnras, 484, 1031,
  \dodoi{10.1093/mnras/sty3420}

\bibitem[{{Perley} {et~al.}(2020){Perley}, {Fremling}, {Sollerman}, {Miller},
  {Dahiwale}, {Sharma}, {Bellm}, {Biswas}, {Brink}, {Bruch}, {De}, {Dekany},
  {Drake}, {Duev}, {Filippenko}, {Gal-Yam}, {Goobar}, {Graham}, {Graham}, {Ho},
  {Irani}, {Kasliwal}, {Kim}, {Kulkarni}, {Mahabal}, {Masci}, {Modak}, {Neill},
  {Nordin}, {Riddle}, {Soumagnac}, {Strotjohann}, {Schulze}, {Taggart},
  {Tzanidakis}, {Walters}, \& {Yan}}]{Perley20}
{Perley}, D.~A., {Fremling}, C., {Sollerman}, J., {et~al.} 2020, \apj, 904, 35,
  \dodoi{10.3847/1538-4357/abbd98}

\bibitem[{{Perley} {et~al.}(2021){Perley}, {Ho}, {Yao}, {Fremling}, {Anderson},
  {Schulze}, {Kumar}, {Anupama}, {Barway}, {Bellm}, {Bhalerao}, {Chen}, {Duev},
  {Galbany}, {Graham}, {Gromadzki}, {Guti{\'e}rrez}, {Ihanec}, {Inserra},
  {Kasliwal}, {Kool}, {Kulkarni}, {Laher}, {Masci}, {Neill}, {Nicholl},
  {Pursiainen}, {van Roestel}, {Sharma}, {Sollerman}, {Walters}, \&
  {Wiseman}}]{Perley21}
{Perley}, D.~A., {Ho}, A. Y.~Q., {Yao}, Y., {et~al.} 2021, \mnras, 508, 5138,
  \dodoi{10.1093/mnras/stab2785}

\bibitem[{{Pursiainen} {et~al.}(2018){Pursiainen}, {Childress}, {Smith},
  {Prajs}, {Sullivan}, {Davis}, {Foley}, {Asorey}, {Calcino}, {Carollo},
  {Curtin}, {D'Andrea}, {Glazebrook}, {Gutierrez}, {Hinton}, {Hoormann},
  {Inserra}, {Kessler}, {King}, {Kuehn}, {Lewis}, {Lidman}, {Macaulay},
  {M{\"o}ller}, {Nichol}, {Sako}, {Sommer}, {Swann}, {Tucker}, {Uddin},
  {Wiseman}, {Zhang}, {Abbott}, {Abdalla}, {Allam}, {Annis}, {Avila}, {Brooks},
  {Buckley-Geer}, {Burke}, {Carnero Rosell}, {Carrasco Kind}, {Carretero},
  {Castander}, {Cunha}, {Davis}, {De Vicente}, {Diehl}, {Doel}, {Eifler},
  {Flaugher}, {Fosalba}, {Frieman}, {Garc{\'\i}a-Bellido}, {Gruen}, {Gruendl},
  {Gutierrez}, {Hartley}, {Hollowood}, {Honscheid}, {James}, {Jeltema},
  {Kuropatkin}, {Li}, {Lima}, {Maia}, {Martini}, {Menanteau}, {Ogando},
  {Plazas}, {Roodman}, {Sanchez}, {Scarpine}, {Schindler}, {Smith},
  {Soares-Santos}, {Sobreira}, {Suchyta}, {Swanson}, {Tarle}, {Tucker},
  {Walker}, \& {DES Collaboration}}]{Pursiainen18}
{Pursiainen}, M., {Childress}, M., {Smith}, M., {et~al.} 2018, \mnras, 481,
  894, \dodoi{10.1093/mnras/sty2309}

\bibitem[{{Raskin} {et~al.}(2012){Raskin}, {Scannapieco}, {Fryer},
  {Rockefeller}, \& {Timmes}}]{2012ApJ...746...62R}
{Raskin}, C., {Scannapieco}, E., {Fryer}, C., {Rockefeller}, G., \& {Timmes},
  F.~X. 2012, \apj, 746, 62, \dodoi{10.1088/0004-637X/746/1/62}

\bibitem[{{Rueda} {et~al.}(2013){Rueda}, {Boshkayev}, {Izzo}, {Ruffini},
  {Lor{\'e}n-Aguilar}, {K{\"u}lebi}, {Aznar-Sigu{\'a}n}, \&
  {Garc{\'{\i}}a-Berro}}]{2013ApJ...772L..24R}
{Rueda}, J.~A., {Boshkayev}, K., {Izzo}, L., {et~al.} 2013, \apjl, 772, L24,
  \dodoi{10.1088/2041-8205/772/2/L24}

\bibitem[{{Rueda} {et~al.}(2018){Rueda}, {Ruffini}, {Wang}, {Aimuratov},
  {Barres de Almeida}, {Bianco}, {Chen}, {Lobato}, {Maia}, {Primorac},
  {Moradi}, \& {Rodriguez}}]{2018JCAP...10..006R1}
{Rueda}, J.~A., {Ruffini}, R., {Wang}, Y., {et~al.} 2018, \jcap, 10, 006,
  \dodoi{10.1088/1475-7516/2018/10/006}

\bibitem[{{Rueda} {et~al.}(2019){Rueda}, {Ruffini}, {Wang}, {Bianco},
  {Blanco-Iglesias}, {Karlica}, {Lor{\'e}n-Aguilar}, {Moradi}, \&
  {Sahakyan}}]{2019JCAP...03..044R}
---. 2019, \jcap, 2019, 044, \dodoi{10.1088/1475-7516/2019/03/044}

\bibitem[{{Ruiter} {et~al.}(2009){Ruiter}, {Belczynski}, \&
  {Fryer}}]{2009ApJ...699.2026R}
{Ruiter}, A.~J., {Belczynski}, K., \& {Fryer}, C. 2009, \apj, 699, 2026,
  \dodoi{10.1088/0004-637X/699/2/2026}

\bibitem[{{Schwab} {et~al.}(2016){Schwab}, {Quataert}, \&
  {Kasen}}]{2016MNRAS.463.3461S}
{Schwab}, J., {Quataert}, E., \& {Kasen}, D. 2016, \mnras, 463, 3461,
  \dodoi{10.1093/mnras/stw2249}

\bibitem[{{Schwab} {et~al.}(2012){Schwab}, {Shen}, {Quataert}, {Dan}, \&
  {Rosswog}}]{2012MNRAS.427..190S}
{Schwab}, J., {Shen}, K.~J., {Quataert}, E., {Dan}, M., \& {Rosswog}, S. 2012,
  \mnras, 427, 190, \dodoi{10.1111/j.1365-2966.2012.21993.x}

\bibitem[{{Shen} {et~al.}(2012){Shen}, {Bildsten}, {Kasen}, \&
  {Quataert}}]{2012ApJ...748...35S}
{Shen}, K.~J., {Bildsten}, L., {Kasen}, D., \& {Quataert}, E. 2012, \apj, 748,
  35, \dodoi{10.1088/0004-637X/748/1/35}

\bibitem[{{Shen} {et~al.}(2019){Shen}, {Quataert}, \& {Pakmor}}]{Shen19}
{Shen}, K.~J., {Quataert}, E., \& {Pakmor}, R. 2019, \apj, 887, 180,
  \dodoi{10.3847/1538-4357/ab5370}

\bibitem[{{Sousa} {et~al.}(2020{\natexlab{a}}){Sousa}, {Coelho}, \& {de
  Araujo}}]{2020MNRAS.492.5949S}
{Sousa}, M.~F., {Coelho}, J.~G., \& {de Araujo}, J. C.~N. 2020{\natexlab{a}},
  Monthly Notices of the Royal Astronomical Society, 492, 5949,
  \dodoi{10.1093/mnras/staa205}

\bibitem[{{Sousa} {et~al.}(2020{\natexlab{b}}){Sousa}, {Coelho}, \& {de
  Araujo}}]{2020MNRAS.498.4426S}
---. 2020{\natexlab{b}}, \mnras, 498, 4426, \dodoi{10.1093/mnras/staa2683}

\bibitem[{{Sousa} {et~al.}(2022){Sousa}, {Coelho}, {de Araujo}, {Kepler}, \&
  {Rueda}}]{2022ApJ...941...28S}
{Sousa}, M.~F., {Coelho}, J.~G., {de Araujo}, J.~C.~N., {Kepler}, S.~O., \&
  {Rueda}, J.~A. 2022, \apj, 941, 28, \dodoi{10.3847/1538-4357/aca015}

\bibitem[{{Stroeer} \& {Vecchio}(2006)}]{2006CQGra..23S.809S}
{Stroeer}, A., \& {Vecchio}, A. 2006, Classical and Quantum Gravity, 23, S809,
  \dodoi{10.1088/0264-9381/23/19/S19}

\bibitem[{{Tampo} {et~al.}(2020){Tampo}, {Tanaka}, {Maeda}, {Yasuda},
  {Tominaga}, {Jiang}, {Moriya}, {Morokuma}, {Suzuki}, {Takahashi}, {Kokubo},
  \& {Kawana}}]{Tampo20}
{Tampo}, Y., {Tanaka}, M., {Maeda}, K., {et~al.} 2020, \apj, 894, 27,
  \dodoi{10.3847/1538-4357/ab7ccc}

\bibitem[{{Tanaka} {et~al.}(2016){Tanaka}, {Tominaga}, {Morokuma}, {Yasuda},
  {Furusawa}, {Baklanov}, {Blinnikov}, {Moriya}, {Doi}, {Jiang}, {Kato},
  {Kikuchi}, {Kuncarayakti}, {Nagao}, {Nomoto}, \& {Taniguchi}}]{Tanaka16}
{Tanaka}, M., {Tominaga}, N., {Morokuma}, T., {et~al.} 2016, \apj, 819, 5,
  \dodoi{10.3847/0004-637X/819/1/5}

\bibitem[{{Tylenda} {et~al.}(2011){Tylenda}, {Hajduk}, {Kami{\'n}ski},
  {Udalski}, {Soszy{\'n}ski}, {Szyma{\'n}ski}, {Kubiak}, {Pietrzy{\'n}ski},
  {Poleski}, {Wyrzykowski}, \& {Ulaczyk}}]{Tylenda11}
{Tylenda}, R., {Hajduk}, M., {Kami{\'n}ski}, T., {et~al.} 2011, \aap, 528,
  A114, \dodoi{10.1051/0004-6361/201016221}

\bibitem[{{Webbink}(1984)}]{1984ApJ...277..355W}
{Webbink}, R.~F. 1984, \apj, 277, 355, \dodoi{10.1086/161701}

\bibitem[{{Zhu} {et~al.}(2013){Zhu}, {Chang}, {van Kerkwijk}, \&
  {Wadsley}}]{2013ApJ...767..164Z}
{Zhu}, C., {Chang}, P., {van Kerkwijk}, M.~H., \& {Wadsley}, J. 2013, \apj,
  767, 164, \dodoi{10.1088/0004-637X/767/2/164}

\end{thebibliography}
\bibliographystyle{aasjournal}

\end{document}